\documentclass[acmsmall]{acmart}
\AtBeginDocument{%
  }

\usepackage[english]{babel}
\usepackage{multirow}
\usepackage{subcaption}

\usepackage{tikz}
\usepackage{amsmath}
\usepackage{xcolor}
\usepackage{xspace}
\usepackage{graphicx}
\usepackage[normalem]{ulem}
\usepackage{caption}
\usepackage{wrapfig}
\usepackage{enumitem}

\newcommand{\sys}{AdvNet\xspace}
\newcommand{\reference}{\texttt{reference}\xspace}
\newcommand{\target}{\texttt{target}\xspace}

\begin{document}
\acmConference
\acmBooktitle
\setcopyright{cc}
\setcctype{by}
\acmJournal{PACMNET}
\acmYear{2026} \acmVolume{4} \acmNumber{CoNEXT2} \acmArticle{12}
\acmMonth{6} \acmDOI{10.1145/3808660}
\received{December 2025}
\received[accepted]{April 2026}
\title{\sys: Revealing Performance Issues in Network Protocols by Generating Adversarial Environments}

\author{Shehab Sarar Ahmed}
\orcid{0000-0002-6063-2689}
\affiliation{%
  \institution{University of Illinois Urbana-Champaign}
  % \department{Computer Science}
  \city{Urbana}
  \state{Illinois}
  \country{USA}
}
\affiliation{%
  \institution{BUET}
  \city{Dhaka}
  % \state{Illinois}
  \country{Bangladesh}
}
\author{William Sentosa}
\orcid{0009-0001-5651-4786}
\affiliation{%
  \institution{University of Illinois Urbana-Champaign}
  % \department{Computer Science}
  \city{Urbana}
  \state{Illinois}
  \country{USA}
}
\author{Yinjie Zhang}
\orcid{0009-0006-0959-5452}
\affiliation{%
  \institution{University of Illinois Urbana-Champaign}
  % \department{Computer Science}
  \city{Urbana}
  \state{Illinois}
  \country{USA}
}
\author{Yoav Lebendiker}
\orcid{0009-0009-7133-9435}
\affiliation{%
  \institution{The Hebrew University of Jerusalem}
  % \department{Computer Science}
  \city{Jerusalem}
  % \state{Illinois}
  \country{Israel}
}
\author{Mickey Shnaiderman}
\orcid{0009-0007-9973-3261}
\affiliation{%
  \institution{The Open University of Israel}
  % \department{Computer Science}
  \city{Raanana}
  % \state{Illinois}
  \country{Israel}
}
\author{Tomer Gilad}
\orcid{0009-0003-5319-4240}
\affiliation{%
  \institution{The Hebrew University of Jerusalem}
  % \department{Computer Science}
  \city{Jerusalem}
  % \state{Illinois}
  \country{Israel}
}
\author{Nathan H. Jay}
\orcid{0009-0003-3139-6220}
\affiliation{%
  \institution{University of Illinois Urbana-Champaign}
  % \department{Computer Science}
  \city{Urbana}
  \state{Illinois}
  \country{USA}
}
\author{Brighten Godfrey}
\orcid{0009-0003-2930-1982}
\affiliation{%
  \institution{University of Illinois Urbana-Champaign}
  % \department{Computer Science}
  \city{Urbana}
  \state{Illinois}
  \country{USA}
}
\author{Michael Schapira}
\orcid{0000-0002-9336-8351}
\affiliation{%
  \institution{The Hebrew University of Jerusalem}
  % \department{Computer Science}
  \city{Jerusalem}
  % \state{Illinois}
  \country{Israel}
}

\renewcommand{\shortauthors}{Shehab et al.}

\begin{abstract}
Infrastructure protocols like Congestion Control (CC) seek to provide reliable performance across a wide range of Internet environments.
Currently, protocol designers assess performance through hand-designed test cases or data sets captured from real environments.
However, such approaches may inadvertently overlook critical facets of the algorithm's behavior when they encounter an unanticipated environment or workload.

We seek to understand the unanticipated with \sys, a system that automatically generates adversarial network environments that cause a target protocol implementation to perform poorly.
\sys employs machine learning-based optimization to generate environments, and incorporates a robust noise-handling mechanism to mitigate the variability inherent in real-world protocol performance.
Although our approach is more general, this paper focuses specifically on transport protocols and their CC implementations.
We showcase \sys's capability to create adversarial scenarios for 27 kernel-space implementations of both single-path and multi-path CC protocols, for several use cases with different performance goals.
\sys identifies problematic network conditions that expose previously unnoticed Linux kernel bugs and uncovers hidden limitations in CC implementations, and provides insights about robustness.
These results suggest that automated adversarial testing can be a valuable tool in protocol development, and that robustness is a useful new dimension for benchmarking CC protocols.
\end{abstract}



\maketitle

\section{Introduction}
\label{sec:intro}

Applications, algorithms, and protocols that depend on the underlying network performance face a challenge in testing: networks are very diverse and encounter many possible conditions.
This is particularly a problem for network protocols and systems like transport protocols that form infrastructure that an enormous number of applications and users depend on, across every corner of the Internet.
We expect these protocols to perform dependably, across a vast range of conditions, encompassing varying network topologies, bandwidths, latencies, loss rates, and traffic loads, and patterns of change across time in all of the above.

As a specific example, imagine a scenario where a protocol designer is tasked with developing an upgraded version of an existing protocol.
The innate intricacies associated with this new version, coupled with the extensive spectrum of environments in which the protocol is intended to function, pose a considerable testing challenge.
The changes might be intended to help performance, but are there network environments where the changes might result in a poor tradeoff?

The most common way of testing a network protocol is running it in a test environment with manually-constructed parameters, e.g., a certain range of fixed bandwidth or latency~\cite{zhang2019evaluation}.  
An expanded set of tests might involve a dataset or trace of performance from a real environment which can be replayed in emulation~\cite{white2002integrated, netravali2015mahimahi,sentosa2023dchannel, wischik2011design}, or even running the protocol in a limited set of deployment tests~\cite{yan2018pantheon}.
However, these methods only cover a minuscule fraction of the diverse environments in which the algorithm might operate.

We propose that an effective way to address these challenges is to automatically discover instances where the protocol under test exhibits unexpected behavior.
To that end, we introduce a simple, yet effective, framework for \emph{automatically generating adversarial environments} for network protocols, with a focus here on congestion control.
This framework, \sys, trains a machine learning-based optimization algorithm whose goal is to create environments that damage performance of the target protocol (hereafter referred to as \target).
An \emph{environment} here is a time-series or trace of network performance parameters across time, including bandwidth and latency, along with constant parameters like buffer size or total bytes to transmit.
\begin{wrapfigure}{r}{0.5\linewidth}
\vspace{-5pt}
  \includegraphics[trim={0cm 3.9cm 0cm 3.9cm},clip,width=\linewidth]{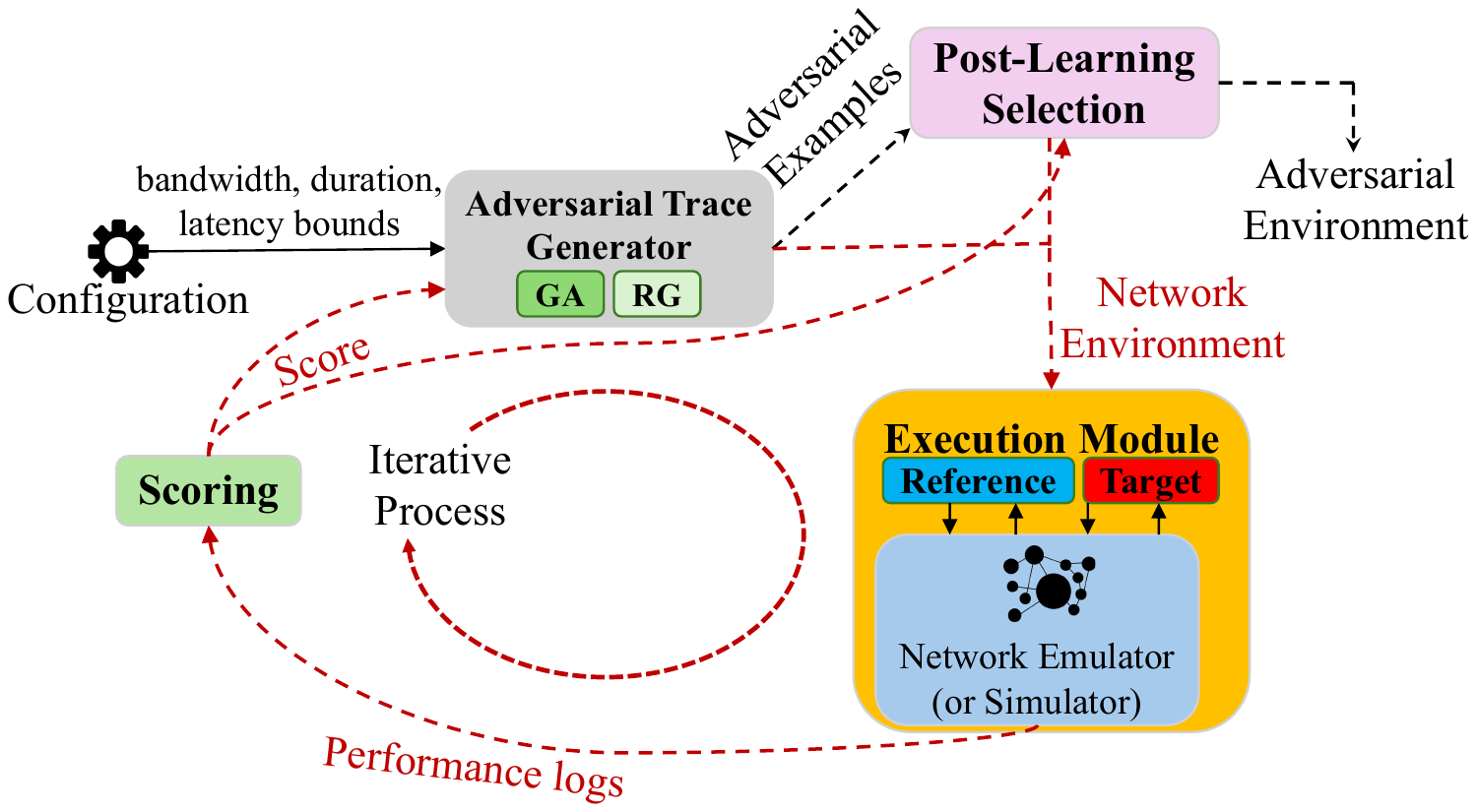}
  \caption{
  High-level architecture of \sys.
  } 
  \label{fig:arch}
  \vspace{-5pt}
\end{wrapfigure}
Intuitively, an adversarial environment is one where \target performs poorly, despite having the potential to improve its performance.
To quantify this, \sys allows the user to specify a reference measure of performance (hereby referred as \reference) that can be either a concrete protocol execution, or a calculation. 
\sys's adversary then attempts to generate environments that maximize the performance gap  between \reference and \target, which we call the \texttt{score}. \sys's high-level architecture is shown in Figure \ref{fig:arch}.

Realizing the above approach brings both algorithmic and systems challenges. \textbf{(1) Sample efficiency:} A key goal of our work is to test real-world production code such as the Linux kernel and user-space implementations of TCP (rather than just simulations of their algorithms). The bottleneck is therefore running emulations of the target protocol in various network traces, meaning the adversary's algorithm must be sample-efficient. Complicating matters further, emulation introduces non-deterministic noise that can confuse learning, especially since combating noise by repeatedly re-running emulations is expensive. \textbf{(2) Practical utility:} A second key goal is to explore whether adversarial testing can be practically useful for congestion control protocol implementations, which requires building a flexible implementation and applying it to use cases.

Adversarial environment generation for CC was previously proposed in a workshop paper that we extend~\cite{gilad2019robustifying} and in a more recent workshop paper, CC-Fuzz~\cite{ray2022cc}. But \cite{gilad2019robustifying,ray2022cc} did not solve the above challenges. CC-Fuzz runs in simulation, and when moving to emulation, we found that some of its design decisions result in relatively sample-inefficient learning in the sense that it achieves a relatively low score in a given budget of trace emulations 
% \cut{Furthermore, while \cite{ray2022cc} discovered several protocol issues in simulation, it did not demonstrate that adversarial testing can evaluate useful properties of real protocol implementations.}
(see \S\ref{sec:related} for a more complete comparison between our work and \cite{gilad2019robustifying,ray2022cc}). Our work addresses these challenges, as follows.

%wheas we seek to test \emph{real protocol implementations}. We found that the past approaches were lacking both because they are relatively \emph{sample-inefficient} which becomes problematic when running a relatively expensive real-protocol emulation, and because emulation introduces \emph{significant variability} that confuses learning approaches not designed to handle it. Second, the idea of adversarial testing was not validated with practical use cases and extensive tests on real protocol implementations. 

%Fleshing out the above approach into a system requires solving several problems.

%We want \sys to be practical and easy to use on production code such as the Linux kernel and user-space implementations of TCP algorithms. Past uses of learning in the context of congestion control (such as learning-based congestion controllers~\cite{abbasloo2020classic, jay2019deep}) are trained in simulated environments. Running real code in emulation makes \sys's learning problem harder due to the noisiness of results, and the fact that fewer training examples can be produced in a given amount of time.

\textbf{Sample-efficient learning algorithms.} We made several design choices to enable effective adversarial testing with a limited number of emulation runs. First, \sys uses a relatively compact environment representation that is less dependent on specific packet timings (unlike CC-Fuzz's \cite{ray2022cc} use of packet delivery opportunity traces). Second, we evaluate several learning-based optimizers, including genetic algorithms (GA), Bayesian optimization (BO), and bandit learning (BL). Finally, we found that simple selection of the most adversarial trace is suboptimal due to emulation variance, and instead design a post-learning selection (PLS) algorithm which makes a better choice.

%making the learning process aware of variance in results to try to steer it towards more stable behaviors and run, 

\textbf{Use cases and associated performance goals.} \sys allows a customizable definition of the \reference, but how to use this differs depending on the use case.
We specialize \sys by defining different realizations of \reference and {\target} for three use cases where adversarial testing is valuable: (UC1) finding suboptimal performance of individual protocols; (UC2) regression testing, i.e., finding cases where a new protocol version degrades performance relative to a prior version; and (UC3) validating specific performance objectives of the protocol designer.

%\textbf{(UC1) Finding Suboptimal Performance:}
% We can calculate the maximum achievable performance, e.g., the available bandwidth summed across time, as \reference.
% Although no congestion control implementation is expected to be perfect, a large gap between \reference and \target may indicate a problem.
%We can set \reference to maximum achievable bandwidth to find sub-optimal performance of TCP implementations. A large gap between \reference and \target may indicate a problem. \textbf{(UC2) Regression testing:} This involves determining whether a new protocol version has introduced a flaw: Here, \reference is an older version of the protocol and \target is the new version. \textbf{(UC3) Validating design objectives:} Here, the protocol designer wishes to test a performance property that should hold.
% : for example, multipath TCP protocols (MPTCP) are explicitly intended~\cite{kato2019experimental} to perform at least as well as when using only a single one of the available paths.
% In that case, \target can be an MPTCP implementation running on two paths (so \sys generates network environments consisting parameters for two parallel network links), and \reference is the same protocol running on just one of the paths.

\textbf{Implementation.} We implemented \sys including the learning module, integration with the Mahimahi~\cite{netravali2015mahimahi} emulator, and orchestration to run tests. We extended Mahimahi to support multipath TCP, time-varying latency, and more efficient bandwidth trace input. \sys logs packet-level events and the associated actions the emulator took on those packets; we found these logs useful to root-cause issues that \sys exposed. We also extended the emulator to support parallel execution, enabling more trials without affecting wall-clock time.

%\textbf{Understanding the results.} Although an adversarial example is already useful to protocol designers, \sys would be more useful if it provides the opportunity to identify the root cause of the adversarial behavior.  To move in this direction, we enhanced the Mahimahi emulator \cite{netravali2015mahimahi} to provide detailed insights into various events related to network packets.  Using these detailed logs generated from the adversarial examples, we identified several anomalous behaviors in Linux kernel TCP implementations. 

\textbf{Evaluation.}
We begin by testing \sys's design choices. First, we show that \sys's representation of the network environment (with time-varying bandwidths and latencies) allows it to learn higher scores, outperforming CC-Fuzz's representation by about 43\% given the same $\approx$ 1 hour compute time (\S\ref{sec:cc_fuzz}).
Second, we show that our PLS algorithm improves performance by 26\% over the baseline (\S\ref{sec:selection}). Third, we compared candidate optimization approaches, showing (among other results) that a Genetic Algorithm (GA) outperforms Random Generation (RG) by 23\% on average (\S\ref{sec:opt_alg}).
% Finally, we found that parallelism provides only a small improvement due to emulation bottlenecks, suggesting an opportunity for future work (\S\ref{sec:parallelism}).

%Third, we analyzed how the number of executions per environment affects the adversarial environments discovered by \sys. We found that even a single evaluation yields a sufficiently strong score. Although increasing the number of executions initially degrades performance due to noise and limited sampling, the performance gradually improves, with five parallel executions providing the best overall results (\S\ref{sec:parallelism}).
%Finally, to address the inherent noise present in real-world protocol performance, we apply a selection phase after the optimization algorithm completes. We evaluated several selection strategies and found that the Multi-Stage Tournament (MST) method consistently performs the best. We further examined how much of the total training time should be allocated to this selection phase and determined that dedicating 10\% of the runtime to MST yields the highest overall performance (\S\ref{sec:selection}).

Having established our method, we applied \sys to the three use cases mentioned above, running an extensive set of tests across 27 transport protocol implementations. The goal of these tests was to determine whether \sys is able to successfully discover adversarial environments and whether those examples are useful. Key results include the following:
\begin{itemize}[noitemsep,topsep=0pt,parsep=0pt,partopsep=0pt,leftmargin=*]
    \item \sys is able to find adversarial environments in most cases we tested -- including 448 of 544 scenarios of two ``all pairs'' tests, with 17 Linux kernel TCPs as the \target vs. every other as the \reference. The magnitude of the adversary's score varies greatly across scenarios, which gives insight into relative protocol robustness and points toward especially serious potential problems.

    \item (UC1): We root-caused several of the adversarial environments that \sys generated, identifying a rare corner-case bug in the Linux kernel that causes TCP to stop sending packets and enter the retransmission timeout (RTO) phase. We also found a second Linux kernel issue (confirmed by a BBR maintainer) where a TCP flow can abruptly decrease its sending rate and not recover. 
    % and informed us that he would provide a patch. \fix{That's an old statement so we should revise it}

    \item (UC2): We performed regression testing of \texttt{bbr} v3 vs. v1 with respect to the goal of fairness to \texttt{cubic}. \sys found it is possible for v3 to be \emph{more aggressive} than v1. We also compared the two versions with respect to throughput. \sys identified scenarios in which \texttt{bbr} v1 achieves more than \emph{five times} the throughput of \texttt{bbr }v3.

    \item (UC3): We tested the design goal that multipath TCPs should never worsen performance by using a second path. \sys found that all four CC implementations in the Linux kernel MPTCP, as well as DChannel~\cite{sentosa2023dchannel}, can violate that property. We investigated two of those scenarios in more detail (MPTCP's \texttt{balia} algorithm and DChannel).
    
    %MPTCP's \texttt{balia} protocol (where delayed ACK causes a problem) and DChannel (where poor interaction with CUBIC running above it causes the problem).

\end{itemize}

\noindent In summary, this paper makes three key contributions:

\begin{itemize}[topsep=0pt, partopsep=0pt, itemsep=0pt, parsep=0pt]
    % \item We define the objective of our approach through the integration of a reference algorithm, showcasing how this innovative perspective markedly improves the detection of scenarios that unveil bugs or highlight performance issues.\ms{This overstates the importance of the reference algorithm. I think that the main contribution is the first automated framework for generating adversarial examples. The second is realizing it.}
    \item We design and implement an automated framework that generates adversarial environments as a means of identifying performance issues in transport protocols.
    \item We conduct extensive experiments with \sys, and perform root-cause analysis of many of the resulting adversarial environments, demonstrating that adversarial environment generation is a useful way of discovering problems in real protocol implementations.
    \item \sys is available open source\footnote{\url{https://github.com/Dariwala/AdvNet}}, with a customizable interface for users to easily test their chosen protocols.
    Because \sys supports testing protocol implementations (rather than just simulations), it can be easily incorporated into research and development workflows.
\end{itemize} 
We believe that measuring protocol robustness with \sys will be a valuable complement to traditional metrics like throughput and latency on known traces.
While this paper and the current \sys implementation focus on congestion control, \sys can in principle be applied to any protocol with a quantifiable environment and performance metrics, and exploring other domains is a promising area of future work (\S\ref{sec:future-work}).

\section{Why Test Adversarially?}

%As discussed in \S\ref{sec:intro}, i

In both research and production, most testing of transport protocol implementations and their congestion control algorithms (CCAs) use a defined set of network conditions, either manually selected or logged from real-world observations. If these environments indeed mimic expected real-world conditions, isn't that sufficient? 

We argue that finding adversarial environments for protocols, independent of whether similar environments currently appear in deployment, is useful and complementary to traditional testing. An overarching reason is that transport protocols are critical infrastructure that need to work dependably in a diverse range of environments. More concretely, we discuss (and later evaluate) three use cases where we expect adversarial testing to be valuable.
% \vspace{-10pt}
\subsection{Use Cases}
\label{sec:use-cases}

\textbf{Finding environments with sub-optimal performance (UC1).}
Transport protocols, and their CCAs, are notoriously difficult to design with consistently high performance across many environments, as evidenced by the long series of papers developing new designs. Even a minor change in the environment can significantly impact an algorithm’s performance; for example, the throughput of a TCP congestion control algorithm (CCA) can be severely reduced by even a slight increase in packet loss rate~\cite{lakshman2000tcp}.  Adversarial testing can help augment traditional testing to discover unexpected environments that are problematic for a protocol, thus helping protocol designers and implementers improve them and create more robust protocols.

But it is difficult for transport protocols to be optimal, especially measured relative to a theoretical limit. Suppose the performance goal is to maximize throughput; if network bandwidth suddenly increases, TCP cannot be expected to realize this and make use of the capacity immediately. Given that the protocol designer expects some imperfection, why is finding suboptimality useful? There are at least two reasons. First, although some suboptimality is to be expected, egregiously low performance can indicate bugs. Second, we can define the objective to be more realistic: rather than finding cases where a target protocol falls short of a theoretical limit, we can compare it to the performance of other protocol implementations, which are feasible by definition.

\textbf{Regression Testing (UC2)} checks whether modifications (such as redesigns, feature enhancements, or bug fixes) introduced new problems relative to the prior version, and is an integral part of production software releases. Transport protocols today are in a very active state of development, both in user-space transports like QUIC~\cite{iyengar2021quic} and also in the kernel. For example, BBR v1 was integrated into the Linux kernel (version 4.9) in December 2016; Google introduced BBR v2 in 2019 to address some of v1's limitations, and v3 in 2024.

Adversarial testing can find environments where version $n$ performs worse than version $n-1$. Beyond the general desire to achieve high performance as in UC1, adversarial examples are useful here for a couple reasons. First, developers are particularly attuned to understanding the effects of changes in production code, because any new change can introduce a new problem. Indeed, more broadly than just transport protocols, management changes are the leading cause of incidents in clouds~\cite{evolveOrDie}. Second, in some cases the goal of a change might be unrelated to performance, so the expectation is that there is no performance degradation, even in rare cases. Finding such cases is difficult and even tiny differences are important. To take an example outside the domain of transport protocols, Meta goes to great lengths to catch performance regressions in services, with a median performance effect of just 0.14\%~\cite{servicelab}! Worst-case testing is a good fit for such a goal.

\textbf{Validating design objectives (UC3).}
Protocol designers often have specific objectives beyond generic performance goals.
For instance, a primary goal of multi-path TCP (MPTCP) is to outperform single-path TCP \cite{kato2019experimental}. A ``do no harm'' property is therefore relevant: an MPTCP should never perform worse with two paths than if the same protocol had been given just one of the two paths. Other examples might include safety properties like reducing rate when the network has  high loss.

% \vspace{-10pt}
\subsection{Design Goals}
\label{sec:design-goals}

The three use cases above illustrate different situations where protocol developers can benefit from adversarial testing. These cases share some common requirements, which guide the design of \sys to provide a broadly useful framework:

\textbf{(a) Automatically tailored to each specific target protocol:} Different transport protocols have different tradeoffs and weaknesses. In all the above cases, we need to tailor the environment towards what is adversarial for a specific protocol, or even a specific protocol version, to find the most adversarial examples. Automating this is a good fit for learning-based algorithms.

\textbf{(b) Running real code:} Problems can arise not only in algorithms, but also in implementations, and understanding the actual code to be deployed gives the greatest assurance.
%We therefore will use emulated networks underneath real running implementations.

\textbf{(c) Optimization efficiency:} The search space of environments is multi-dimensional (e.g., latency and bandwidth) at even one moment, and problems often arise because of patterns of change across time. Combined with the fact that running real code is slow (relative to simulation), we need algorithms that optimize the adversary's goal in limited time.

\textbf{(d) Flexible objectives:} The above use cases have a general commonality of comparing the performance of a target protocol to some reference (either an ``oracle'' definition of the optimal, or another protocol). However, the specific formulations differ, so the performance objective should be conveniently customizable.

%Inspired by this need, we developed \sys, a framework designed to automatically search for adversarial environments for a target system, helping to identify any corner-cases and potentially improve the system.

%Common approaches to testing networking algorithms include manually configuring environment parameters to create test cases \cite{zhang2019evaluation, ha2008cubic, mo1999analysis, kato2019experimental, ro2016performance}, utilizing existing datasets or collecting network traces from the live internet \cite{white2002integrated, netravali2015mahimahi, sentosa2023dchannel, wischik2011design, mascolo2004performance}, and executing the algorithm in a limited set of deployment environments \cite{yan2018pantheon}.

% \vspace{-10pt}
\section{Design}
The \sys architecture is shown in Figure \ref{fig:arch}.
Within the \textit{Configuration} module, users define constraints (e.g., bounds on the network parameters), select \target and \reference to execute, and set hyperparameters.
The \textit{Adversarial Trace Generator}, or just the Adversary, generates candidate adversarial environments within those constraints. It sends each environment, the \reference, and \target to the \emph{Execution} module
The \emph{Execution} module executes and computes the performance of \reference and \target within a network link created by a \textit{network emulator} (or simulator) using parameters from the network environment.
The \emph{Scoring} module then computes the \texttt{score} of the network environment from the performance logs generated by the Execution module. The \texttt{scores} are given to the Adversary, which iteratively learns from them to attempt to generate more challenging network environments that maximize \texttt{score}. After a set amount of learning time, \emph{Post-Learning Selection (PLS)} uses another limited execution time budget to choose the most adversarial trace found by the Adversary. Next, we discuss each of these components in more detail.

%We begin with the definition of a network environment (or trace), followed by the Adversary and PLS which are the algorithmic components responsible for finding high-score traces. 

% \vspace{-10pt}
\subsection{Network Environment}\label{sec:net_env}

For concreteness we begin by defining the network environment, which is the type of object that \sys will generate.

Our network environment (or trace) is defined by an $N$-dimensional vector consisting of all parameters required to define the network and run \target and \reference within the network. The environment has a series of one or more time intervals. Each time interval has an associated bandwidth, latency, and duration.
Additionally, there are time-invariant parameters: buffer length, and optionally, the amount of data to transmit\footnote{This could be time-variant too representing varying application demands. However, in this version of \sys we assume that either (a) the application is ready to send the full specified amount of data at the beginning of the trace, or (b) if the amount of data parameter is omitted, then we have the emulator send data continuously from start to end of the trace.}. Thus, we can define the network environment as:
$
    T=(B, L, D, I)
$.
Here, $B$ is the vector of bandwidths, $L$ is the vector of latencies,
% Hence, these two sets have the same length.
and $D$ is the vector of durations.
% it might have equal or half of the length of $B$ or $L$ depending on the number of links.
Finally, $I$ is the set of all time-invariant parameters of the network environment. In multipath experiments, the network consists of two links, and therefore each of the time-variant parameters is specified per link, except for the duration (which defines a common time interval for both links).
It is evident that the search space of $T$ is \emph{exponential}.

The input to \sys includes bounds on allowed values for all the above parameters. Specifically, we provide a lower-bound vector $LB=\{l_i:1 \le i \le N\}$ and an upper-bound vector $UB=\{u_i:1 \le i \le N\}$ to \sys.
The optimization algorithm running within \sys ensures that each network environment $T=\{t_i:1 \le i \le N\}$, satisfies
$l_i \le t_i \le u_i$.

% \vspace{-10pt}
\subsection{Adversary}\label{sec:adv_trace}

\subsubsection{Adversary design choices}

\textbf{Trace-based vs. action-based adversaries.} A trace-based adversary produces a complete trace, which is a time-ordered sequence of network conditions such as bandwidth and latency, as a single output.
This type of adversary has the advantage of having access to the entire trace, enabling it to manipulate different aspects of the trace and observe the immediate effects of those changes.
It can efficiently determine which parts of the trace to modify and in what direction to create more effective traces.
However, this advantage may turn into a disadvantage when dealing with highly complex (high-dimensional) traces.

On the other hand, an action-based adversary engages with a protocol by taking a series of actions, closely observing the protocol's behavior at a fine-grained level, and providing only the next network conditions.
For example, it might generate a single set of values representing bandwidth and latency for a specific duration.
This type of adversary is adept at learning to create partial traces that lead to poor performance of \target during that specific duration.
However, integrating a protocol's execution with an action-based adversary requires additional engineering complexity, as it necessitates monitoring and influencing the protocol's live behavior and syncing the execution of \reference and \target, which would make it harder for users of \sys to easily extend the framework to test different protocols.
% Additionally, because we want to execute real code in emulation (rather than in a simulator), the action-based adversary must make decisions in real-time, which could limit its ability to take effective actions (e.g., because of limited compute time).
% Thus, it is important to introduce a separation between the adversary and protocol execution.
We therefore opted to work with \emph{trace-based} adversaries.

\textbf{Measuring success of the adversary.} At a high level, we want the adversary to find environments when the target algorithm performs poorly.
But that statement is too simplistic: the adversary could simply design an environment which has zero throughput or 100\% packet loss, which certainly results in poor performance but is uninteresting.
Instead, we need to measure the adversary's success relative to some sort of reference point.

We thus incorporate a \emph{reference} measurement of performance (\reference) into our framework.
We instruct the adversary to generate environments that maximize the performance gap between \reference and \target.
When a substantial performance gap emerges, it serves as an automatic indicator that the environment has successfully caused \target to underperform.
% One option for \referen
The \textit{Adversary} generates network environments and sends them to the Execution module which creates network link(s) based on the environment's parameters (e.g., time-varying bandwidths, latencies).
We then direct the \emph{Execution} module to assess \target and \reference within the network link(s).
The relative difference in performance scores between \reference and \target, as provided by the \emph{Scoring} module, serves as feedback to the optimization algorithm.
This feedback loop helps the optimization algorithm learn to generate environments that result in a larger performance difference between \reference and \target.

% \fix{Redundancy. Can delete/move some of this to \S\ref{subsec:objective}.}The selection of \reference for the learning process, multiple options are available.
% One approach is to use a well-established and trusted algorithm as \reference.
% Another option is to select an earlier version of \target.
% Alternatively, if the domain permits easy calculation of optimal performance, this can serve as \reference. Both \reference and \target depend on the use case and are customizable; we discuss 
% these in more detail in \S\ref{subsec:objective}.

\textbf{Stopping criterion.} In general, the learning phase of the adversary will improve with increased runtime. In our framework, we run the learning phase for a fixed duration or a predetermined number of iterations, which provides a consistent basis for comparing different adversarial strategies.

%A crucial consideration during the learning phase of the adversary is determining when to halt the learning process.
% Commonly, optimization algorithms can become trapped in local optima.
% Two viable options for addressing this issue are as follows:
% \textbf{(1) Monitoring improvement:} Continuously monitor the improvement in the achieved score over time, and if no significant improvement occurs for a substantial duration, terminate the training process.
% \textbf{(2) Fixed time or number of iterations:} Train the adversary for a predefined, fixed time or number of iterations, and report the highest score achieved within that time frame.
% This approach simplifies the training process and facilitates a fair comparison between different adversaries.

% \vspace{-5pt}
\subsubsection{Optimization Algorithms}
% \label{sec:opt_alg}
% We initially selected GA as the optimization algorithm. Although \textit{1-step reinforcement learning} could also be a promising deep-learning approach to explore, we did not pursue deep-learning techniques, as GA yielded satisfactory results in our experiments.

% \textcolor{red}{
There are several candidate optimization algorithms suitable for training \sys's adversary, including, but not limited to, evolutionary algorithms, Bayesian Optimization (BO), Bandit Learning (BL), and Reinforcement Learning (RL).
% }

% Bayesian Optimization (BO) learns a probabilistic model to predict the value of an unknown function at points where it has not yet been evaluated. However, we found that this approach did not perform well with real-life algorithms, where performance can be highly unpredictable.

% We also explored Bandit Learning (BL) using different strategies such as $\epsilon$-greedy, Upper-Confidence Bound, and Thompson Sampling. Unfortunately, BL also delivered poor results, and we found no viable alternative to a neural network for learning an effective policy. 

Although RL appears to be a natural choice, its effectiveness relies on selecting actions online while both \reference and \target are executing. In practice, this requires tight synchronization between the parallel runs of \reference and \target as well as the adversary. We found this synchronization requirement difficult to satisfy reliably. Small timing differences, scheduling latency, and inherent nondeterminism in executions often caused the two protocols to diverge in state, making it unclear which action was being applied to which underlying condition. As a result, RL agents frequently operated on stale or mismatched information, causing learning instability and ineffective policy updates. While these issues are likely solvable, the synchronization requirement introduces complexity and impacts overall flexibility of the design.
% As a result, we switched to 1-step RL, which involves generating the entire environment at once and receiving the reward. For brevity, we will refer to this as RL from this point forward.

% \textcolor{red}{
Therefore, we excluded RL and selected a diverse set of optimization strategies: GA from evolutionary methods, $\epsilon$-greedy search (EPS) as a lightweight variant of BL, BO as a model-based approach, and Random Generation (RG) as a baseline.
% } 

\textbf{Genetic Algorithm (GA).}
The idea behind GA is to create a population of potential \emph{individuals} to a problem, evaluate their fitness based on a defined objective function, and then apply operators like selection, crossover (recombination), and mutation to create new generations of individuals.
These new generations are iteratively refined to improve the fitness of the solutions.

One limitation of the network emulator we use is that it only accepts integer values.
Therefore, it is essential that mutation and crossover operations do not introduce floating-point numbers (rounding is an option, but would create confusion for the optimization algorithm), as this would alter the integrity of the traces.
To ensure this, we employed 2-point crossover \cite{spears1991analysis} and uniform mutation. Below is a brief description of these methods:

\textbf{(1) 2-point crossover:} Two random points are selected within the parent environments. The segments between these points are swapped between the parents to generate new offspring.

\textbf{(2) Uniform mutation:} For each element $t_i$ in the trace, there is a predefined probability \emph{mut\_prob} that $t_i$ will be replaced by a randomly selected integer between its corresponding lower and upper bounds $l_i$ and $u_i$ (\S\ref{sec:net_env} describes the notations).

% \textcolor{red}{
\textbf{Bayesian Optimization (BO).}
BO is a sample-efficient global optimization technique that models the objective function using a probabilistic surrogate and selects new evaluation points by balancing exploration and exploitation. In our implementation, we use a tree-based surrogate model (random forest) to approximate the mapping from network environments to their corresponding \emph{score}.
% } 

\textcolor{black}{
At each iteration, the surrogate model is updated with previously evaluated samples, and an acquisition function is used to determine the next candidate environment. Specifically, we employ the Lower Confidence Bound (LCB) acquisition function, which favors points with either high predicted score or high uncertainty. This enables systematic exploration of the search space while progressively focusing on promising regions.}

% Similar to the GA setting, the network emulator only accepts integer-valued inputs. Therefore, the search space is defined over integer bounds for each trace element $t_i \in [l_i, u_i]$, ensuring that all generated candidates remain valid without requiring post-processing such as rounding.

% We initialize the optimization with randomly sampled points, and iteratively evaluate new candidates until the evaluation budget is exhausted. To ensure consistency with our GA-based experiments, we use the same evaluation function and track the best observed score across iterations.

% \begin{description}
%     \item[2-point crossover:] Two random points are selected within the parent environments. The segments between these points are swapped between the parents to generate new offspring.
%     \item[Uniform mutation:] For each element $t_i$ in the trace, there is a predefined probability \emph{mut\_prob} that $t_i$ will be replaced by a randomly selected integer between its corresponding lower and upper bounds $l_i$ and $u_i$.
% \end{description}
\textcolor{black}{\textbf{$\epsilon$-Greedy Search (EPS).}
We also consider EPS that balances exploration and exploitation without maintaining an explicit population or surrogate model. This approach can be viewed as a variant of multi-armed bandit learning, where candidate network environments correspond to arms and the observed score serves as the reward signal.}

\textcolor{black}{At each iteration, the algorithm selects between two actions: (i) sampling a new trace uniformly at random (exploration), or (ii) refining previously high-performing environments via mutation (exploitation). Specifically, with probability $\epsilon$, a candidate trace is generated by uniformly sampling each element $t_i$ from its corresponding integer range $[l_i, u_i]$. Otherwise, a trace is selected from an \emph{elite set} containing the best-performing candidates observed so far, and a mutation operator is applied. The mutation independently perturbs each element $t_i$ with a fixed probability by replacing it with a randomly sampled integer within its valid bounds.}

\textcolor{black}{To guide the search, we maintain a bounded elite set that stores a subset of the highest-scoring traces encountered during the optimization. This enables the algorithm to exploit promising regions of the search space while retaining diversity through random exploration.} 

% Similar to GA and BO, all candidate traces are constrained to integer values to ensure compatibility with the network emulator. The algorithm iteratively evaluates candidate traces using the same objective function and tracks the best observed score until the evaluation budget or time limit is reached.
\textbf{Random Generation (RG).} Under RG, environments are repeatedly generated by sampling parameter values uniformly from their allowed ranges. This process continues until the evaluation time limit expires. We use RG as a baseline.

\subsection{Post-Learning Selection (PLS)}
\label{sec:pls}

In the process of running the adversary for a set period of time, it generates a large number of environments (e.g., hundreds). The simplest final step would be to return the environment that showed the maximum score. That simple rule, which we call SimpleMax, was our initial implementation. However, we found SimpleMax produces a suboptimal result; and furthermore, when the environment picked by SimpleMax is independently re-tested, it on average produces a significantly lower score than the adversary observed during the training process.

The reason is that execution of an environment in emulation, and the protocols under test themselves, are nondeterministic and variable, with events potentially affected by small timing differences.
When we execute an environment, we will sometimes observe a score that is higher than its true mean score, and sometimes lower. The true mean is not directly observable. By taking the environment whose observed scores were largest, there is a high chance that we are selecting the one which happened to have higher-than-average outliers in the observed scores.

To mitigate this issue, we introduced the \emph{PLS} module. The idea is to spend some time performing a more careful evaluation of the environment that \sys should ultimately output. We dedicate a percentage of \sys's overall execution time budget to PLS; this takes time away from the adversary's learning phase, so PLS needs to make judicious choices in its evaluations. After the adversary (e.g., GA) runs, we retain the top $N$ environments based on their observed scores so far, and these form the input to PLS. PLS has a fixed time budget and has to decide how to distribute that among the $N$ candidate environments. After spending its budget, it then outputs the trace with highest observed mean score.

One might allocate the budget equally among the $N$ environments, which we call RoundRobin. But such uniform allocation is suboptimal. It turns out the problem is more subtle, and has arisen previously in the context of optimizing designs where simulation is expensive. In particular, the Optimal Computing Budget Allocation (OCBA) \cite{478499} seeks to maximize the probability of selecting the candidate with true largest mean. OCBA allocates evaluations adaptively by prioritizing environments exhibiting higher uncertainty or those that are close in score to the current best. Thus, environments with smaller observed performance gaps from the best or larger empirical variances receive more evaluations, leading to statistically efficient identification of the best environment.

However, for our objective of maximizing the mean score of the returned environment, it is possible to do somewhat better. We designed a simple \textbf{Multi-Round Elimination (MRE)} algorithm. Each environment is evaluated once; the half of the candidates with lowest observed mean thus far is discarded. This continues, focusing evaluations on increasingly promising environments. When $\leq5$ environments remain, they are repeatedly evaluated without further elimination, until the evaluation budget is spent. We compared performance of RoundRobin, OCBA, MRE, and other algorithms in a simulator, abstracting the score emulation as sampling Gaussian random variables, and found MRE achieved highest performance (Appendix \ref{appendix:selection_algorithms}), so we use it in our experiments with \sys. It is possible that this algorithm could be improved further in the future~\cite{gao2017new}.

%An important challenge in evaluating adversarial environments is the inherent variance in performance outcomes. This variance arises from randomized effects within the execution environment and nondeterminism introduced by timing-sensitive events.

% total training time in generating adversarial environments and after that, we retain the top N based on their preliminary scores. 

%To mitigate this issue, we introduce the PLS module that operates over the top-ranked environments. Specifically, we dedicate a percentage of total training time in generating adversarial environments and after that, we retain the top N based on their preliminary scores. Each of these environments is subsequently re-evaluated using a selection algorithm. The environment with the highest stable score is then selected. This procedure reduces sensitivity to execution-level randomness and increases confidence that the chosen environment reflects truly adversarial conditions rather than stochastic variance. We explore with the following selection algorithms:

% \vspace{-10pt}
\subsection{Execution Module}
\label{sec:execution-module}
The adversary provides each network environment, along with \reference and \target, to the Execution module. The Execution module parses the network environment and extracts relevant information (e.g., parsing time-varying bandwidths to create uplink and downlink bandwidth files for the network emulator). Based on this information, the network emulator constructs network link(s) according to the specified parameters from the Execution module. The Execution module then runs \reference and \target on the network link(s), and creates their performance logs (e.g., per packet delays, timestamps when packets are sent and received etc.).

%\vspace{-10pt}
%\subsection{Network Simulator vs. Emulator}

\textbf{Emulator.}
We designed \sys with modular components, allowing each part to be modified independently without disrupting the overall workflow. As a result, \sys is compatible with both simulators and emulators, and we have used it with the ns3 simulator. However, we focus on emulation here, to test real protocol implementations.

%\subsubsection{Network Emulator}\label{sec:net_emu}
We enhanced the Mahimahi network emulator \cite{netravali2015mahimahi} to meet our need of emulating time-varying latency and bandwidth.
While Mahimahi (\texttt{mm-link}) natively supports time-varying bandwidth, it does not provide a mechanism for time-varying one-way delay. We extend the emulator to support this functionality by allowing it to ingest a latency trace that specifies a time series of delay values, along with separate bandwidth traces for the uplink and downlink.
% We further modify Mahimahi to enable spawning of multiple emulator instances in parallel. 
% To reduce interference among these parallel executions, we allocate a dedicated CPU core to each emulator process as well as to the client and server processes. To reduce processing overhead, we replace the pre-expanded PDO-style traces with compact high-level bandwidth descriptions, and then compute packet delivery opportunities on demand within the emulator. 
Finally, we extend the emulator from a single-path configuration to a dual-path setting by integrating \texttt{mp-shell}~\cite{wifi-lte-imc14}, enabling us to evaluate protocols such as MPTCP under repeatable and controlled multi-homing environments.

%life algorithms, which would be impossible with simulators.
%For example, we evaluate the Linux kernel implementations of TCP and MPTCP algorithms.
%That said, using an emulator introduces the following challenges:
%\begin{itemize}[noitemsep,topsep=0pt,parsep=0pt,partopsep=0pt,leftmargin=*]
%\item \textbf{Increased Execution Time:} Obtaining \texttt{score} for each network environment takes significantly longer time compared to using a network simulator.
%\item \textbf{Noise:} System uncertainties such as scheduling and interrupts mean that multiple executions of the same protocol do not produce identical results.
%This introduces inherent noise in the performance scores returned by the execution module.
%\end{itemize}
%We discuss how we tackle these issues in the next two sections.

%\label{sec:design-parallelization}
\textbf{Parallelization.}
Training an adversary against real-world protocols presents a key challenge due to the high execution time of protocol runs. Obtaining statistically meaningful estimates of performance requires multiple repeated executions per environment, yet doing so significantly limits the number of environments that can be explored within a practical time budget. To address this scalability bottleneck, we extend Mahimahi to support parallel execution by spawning multiple independent emulator instances. We allocate a dedicated CPU core to each emulator instance, and additionally assign separate cores to the protocol client and server processes to ensure isolation. As a result, evaluating a single environment for one protocol requires three CPU cores operating concurrently. This design allows an environment to be evaluated multiple times simultaneously, effectively reducing wall-clock time and enabling broader exploration of the adversarial space.

% \vspace{-10pt}
\subsection{Computing \texttt{score}}\label{subsec:objective}

As previously mentioned, \sys aims to maximize the performance gap between \reference and \target. Consequently, \texttt{score} should quantify the performance difference of \target relative to \reference.
% While there are various methods to achieve this, the most suitable approach may vary depending on the specific context in which \target is being evaluated. 
% We explore a couple of methods utilized in our experiments:
% We generally use the relative performance difference between \reference and \target as the \texttt{score}.
Depending on how we define algorithm performance, a higher score can be either good or bad. For instance, higher throughput is desirable, while a higher completion time is undesirable.
% If a higher value of the metric is beneficial, we calculate the relative difference between the scores of \reference and \target. Conversely, if a higher value is detrimental, we reverse the calculation to reflect this. 
Formally, \texttt{score} of a network environment is defined as:
\vspace{-8pt}
\begin{equation}\label{eq:first}
    score = 
\begin{cases} 
\frac{\text{reference\_score} - \text{target\_score}}{\text{max(reference\_score, target\_score)}} & \text{if higher score is better} \\
\frac{\text{target\_score} - \text{reference\_score}}{\text{max(reference\_score, target\_score)}} & \text{if lower score is better}
\end{cases}
\end{equation}
\vspace{-25pt}
% We applied this method to compare the pairwise performance of TCP algorithms and to evaluate the performance of MPTCP algorithms relative to SPTCP algorithms.

% The second method considers the relative weight of \reference's performance with respect to \target (with an additional term depending on the context), defined as:
% \begin{equation}\label{eq:second}
% \text{score} = \frac{\text{target\_score}}{\text{reference\_score} + \text{target\_score}}+f(\text{reference}, \text{target})
% \end{equation}
% We applied this method to analyze the fairness of TCP algorithms.

The method for computing reference\_score and target\_score is customizable, depending on the use case. \textcolor{black}{Our experiments used the following:}
\begin{itemize}[noitemsep,topsep=0pt,parsep=0pt,partopsep=0pt,leftmargin=*]
    \item (UC1) We conduct two types of experiments under this use case.
    First, we analyze the sub-optimality of TCP protocols with respect to maximum throughput, by assigning the available bandwidth as reference\_score and the throughput of \target as the target\_score.
    % We compute the available bandwidth based on the environment specification.
    % For computing throughput of \target, we sum the sizes of received packets and divide by the total run time:
    % $
        % \text{throughput} (\tau) = \frac{\sum_{i=1}^{M}p_i}{T}
    % $.
    % Here, $T$ is the total runtime, $M$ is total number of received packets, and $p_i$ is the size of the $i^{th}$ received packet.
    Second, we analyze how unfair a TCP protocol can be to another TCP. In this experiment, 
    % we run both \reference and \target simultaneously within the network environment. We also run \reference separately to ensure the unfairness experienced by \reference is due to the presence of \target. 
    we set reference\_score to be the throughput of \reference when run alone ($\tau_\textrm{ref\_only}$), and set target\_score to be the throughput of \reference when it is run with \target simultaneously ($\tau_\textrm{ref}$).
    \item (UC2) We also do two types of experiments in this use case.
    First, we compare the TCP protocol versions against each other with respect to a weighted sum of average packet delay and throughput. 
    % The throughput is computed according to Equation \ref{eq:tau}.
    % Average packet-delay is computed based on taking the average of all the received packets' one-way delay. We receive the one-way delay of each packet from the \sys's generated log where we have the timestamps for the sending and receival of each packet.
    \textcolor{black}{We compute both reference\_score and target\_score as:
    $
        t_{\textrm{coeff}}\times\tau_{\textrm{rel}}+(1-t_{\textrm{coeff}})\times d_{\textrm{rel}}
    $,
    where $t_{\textrm{coeff}} \in [0,1]$. $\tau_{\textrm{rel}} = \frac{\tau}{\tau_{\max}}$, and $d_{\textrm{rel}} = \frac{d_{\min}}{d}$ denote the relative throughput and relative delay, respectively.
    Here, $\tau$ and $d$ represent the achieved throughput and delay of the protocol.}
    % is a weighting parameter that determines the relative importance of throughput versus latency. 
    % Higher values of $t_{coeff}$ emphasize throughput performance, while lower values place greater emphasis on latency.   
    Second, we study how \target can induce greater unfairness relative to \reference when operating alongside a competing flow. To quantify this effect, we define \texttt{reference\_score} and \texttt{target\_score} as the throughput achieved by the competing flow when run concurrently with \reference and \target, respectively. 
    % Formally, we compute $\tau_{\mathrm{comp}|\mathrm{reference}}$ and $\tau_{\mathrm{comp}|\mathrm{target}}$, where $\tau_{a|b}$ denotes the throughput of flow $a$ when executed alongside flow $b$.
    \item (UC3) To validate MPTCP protocol performance, we calculate the throughput of MPTCP when it is limited to a single link, which we define as the \texttt{reference\_score}, and when it is provided with two links, which we define as the \texttt{target\_score}.
    % and used Equation \ref{eq:first} to compute \texttt{score}.
    Similarly, we evaluated DChannel \cite{sentosa2023dchannel} 
    % a system designed to improve application performance by augmenting a high-bandwidth channel (e.g., 5G eMBB) with a low-latency channel (e.g., 5G URLLC). 
    % By steering packets between the two channels, DChannel aims to surpass the performance of using the high-bandwidth channel alone.
    % Hence, we
    by defining \reference to use only the high-bandwidth channel and \target to use an additional low-latency channel. 
    % and use DChannel to steer packets among the two links.
    % In this experiment, the network environment includes an additional parameter indicating the size of the data transfer. 
    We use flow completion time (FCT) as the performance metric.
    % , and both \texttt{reference\_score} and \texttt{target\_score} are computed from their corresponding FCT measurements.
\end{itemize}

% This would have been the end of the story if performance measurements were noise-free.
% Noise in the execution of real-world protocols is unavoidable, leading to variability across repeated runs of the same environment.
\textcolor{black}{Although we illustrate each use case using CC protocols, we can apply Equation~\ref{eq:first} to any domain where the goal is to maximize the performance gap between \reference and \target.}

To mitigate the issue of noise, while running the learning algorithm, we execute each protocol multiple times per environment and use the median \texttt{reference\_score} and median \texttt{target\_score} across runs.
Furthermore, after \sys outputs the adversarial environment, we do an independent reevaluation of the environment to get rid of any possible bias in reported scores.

% The objective of the adversary is to search for network environments that produce the maximum \texttt{score}, which translates into the greatest performance degradation of \target compared to \reference.

% \vspace{-10pt}
\section{Experimental Results}

\subsection{Methodologies}

% \vspace{-5pt}
\subsubsection{Kernel-space Protocols}
We used Linux kernel version 5.15.0-117-generic for all experiments involving TCP protocols and version 5.4.230.mptcp for all experiments involving MPTCP protocols. Additionally, we used Linux kernel version~6.13.7+ obtained from the official GitHub repository maintained by the Google BBR development team to compare \texttt{bbr} v1 with \texttt{bbr} v3.

% \vspace{-5pt}
% \subsubsection{User-space Protocols}
% We used picoquic for the user-space TCP implementations.
% We developed a simple server and client that use the Picoquic-provided APIs to transfer data via Picoquic's implementations of CC protocols.
\subsubsection{Sending Traffic}
We use iperf to send traffic over the network. The adversary determines the duration of the experiment, and we run iperf for that period. (The experiment with DChannel is slightly different, for historical reasons: the adversary specified the amount of data to transmit, and we run iperf until that amount is sent.)

% \vspace{-5pt}
\subsubsection{Parameter Bounds}
We use bounds for network parameters shown in Table \ref{tab:params_tcp}, except for DChannel which is a completely different experiment and uses bounds in Table~\ref{tab:params_dc} (Appendix~\ref{appendix:dchannel}). We enforce bandwidth at \emph{millisecond granularity}.

%, ensuring that the average bandwidth over any millisecond is approximately equal.

\begin{table*}[h!]
    \centering
    {\small
    \begin{tabular}{c|c|c|c|c|}
    \cline{2-5}
    % &\multicolumn{4}{c|}{\textbf{TCP \& MPTCP}} \\ \cline{2-5}
    &\textbf{Bandwidth}&\textbf{Latency}&\textbf{Duration}&\textbf{Queue Length}\\ \cline{1-5}
    \multicolumn{1}{|c|}{\textbf{Lower Bound}}&1 Mbps&5 ms&0.5 s&500\\ \cline{1-5}
    \multicolumn{1}{|c|}{\textbf{Upper Bound}}&100 Mbps&100 ms&2.5 s&10000\\ \cline{1-5}
    \end{tabular}
    }
    \caption{Parameters used in all experiments with \sys involving TCP and MPTCP protocols.
    The lower and upper bounds for each parameter are defined per timestep.
    % Therefore, if \sys is searching for network environments with 2 timesteps, the lower and upper bounds for each of the two bandwidths in each environment will be 1 Mbps and 1200 Mbps, respectively.
    }
    % \vspace{-5pt}
    \label{tab:params_tcp}
\end{table*}

\vspace{-35pt}
\subsection{Evaluation of \sys Design Choices}
In this section, we evaluate the different components of the design of \sys.
% \vspace{-5pt}
\subsubsection{Environment Definition and Comparison with CC-Fuzz Approach}\label{sec:cc_fuzz}
CC-Fuzz \cite{ray2022cc} utilized fuzzing to create adversarial environments for TCP algorithms implemented in NS3.
There are numerous differences with CC-Fuzz; for a full discussion, see \S\ref{sec:related}. Here we explore one key design difference: the representation of network environment. \sys models the network environment as a dynamic system with varying bandwidth and latency over time, in a defined number of intervals, plus a constant buffer size. In contrast, CC-Fuzz represents the network environment through packet delivery opportunities (PDOs), where each PDO is a specific time instance when the network link can transmit a packet from the sender's queue. Since the CC-Fuzz codebase is not available, we implemented this environment representation approach within \sys.

\begin{wraptable}{r}{0.4\linewidth}
    % \vspace{-2pt}
    \centering
    \begin{tabular}{|c|c|c|}
        \hline
        \textbf{TCP Protocol} & \textbf{\sys} & \textbf{CC-Fuzz} \\
        \hline
        \texttt{bbr}   & 79.11\% & 14.31\% \\
        \hline
        \texttt{cubic}& 46.24\% & 29.78\% \\
        \hline
        \texttt{vegas}& 84.71\% & 27.29\% \\
        \hline
        \texttt{reno} & 41.26\% & 28.8\%  \\
        \hline
    \end{tabular}
    \caption{Best \texttt{score}s achieved by \sys and CC-Fuzz for \textbf{UC1}.}
    \label{tab:advnet_ccfuzz}
    \vspace{-15pt}
\end{wraptable}

To evaluate these two approaches, we conducted experiments using GA as the optimization algorithm with both \sys and CC-Fuzz's PDO approach, keeping all configurations (parameter bounds, runtime, etc.) identical except for the network environment definition.
In this context, \texttt{score} represents the relative difference between the maximum achievable throughput and the throughput of \target. 
Thus, a higher \texttt{score} indicates a more adversarial environment.
For CC-Fuzz's approach, we conducted experiments with four different values for the number of PDOs: 25, 50, 75, and 100.
The best \texttt{score}s for \textbf{UC1} across various TCP protocols are presented in Table \ref{tab:advnet_ccfuzz}.

As is evident, \sys significantly outperforms CC-Fuzz's PDO approach. 
There are two primary reasons for this.  
First, the search space dimensionality of PDO-based representation is inherently higher than that of \sys's time-varying bandwidth and latency representation, leading to a more complex search problem.
We demonstrated this by evaluating what happens when we increase the dimensionality, i.e., the number of PDOs.
Higher dimensionality results in lower \texttt{score} for CC-Fuzz's approach, e.g., 14.31\% with 25 PDOs vs. 5.91\% with 100 PDOs for \texttt{bbr}.
This can be attributed to the fact that as the search space expands, GA struggles to find the optimal solution within the fixed execution time.
Second, by using PDOs, CC-Fuzz does not directly incorporate latency in its definition of the network environment (although PDOs can capture certain latency effects).
As we will discuss later, variations in link latency are critical for generating adversarial scenarios.
Consequently, CC-Fuzz's performance scores are significantly lower compared to those achieved by \sys.

\textcolor{black}{Nevertheless, we emphasize that this comparison isolates only the differences in network environment representation between CC-Fuzz and \sys. The two systems differ along several other dimensions. For instance, CC-Fuzz employs an island-based genetic algorithm, whereas \sys uses a standard genetic algorithm. In addition, CC-Fuzz evaluates both link fuzzing (similar to \sys) and traffic fuzzing, which generates adversarial cross traffic. }

\subsubsection{\textcolor{black}{Post-Learning Selection (PLS)}}\label{sec:selection}
Through our simulation study, we identified MRE as the most effective selection algorithm to apply once optimization algorithm completes (see Appendix~\ref{appendix:selection_algorithms}). After the optimization phase finishes, we take the top 25 discovered environments and pass them to MRE to determine the best environment.

To determine how the runtime should be divided between the optimization and PLS phases, we perform an experiment using \texttt{cubic} as \reference, varying \target over \{\texttt{bbr}, \texttt{highspeed}, \texttt{reno}, \texttt{vegas}\} and GA as the optimization algorithm. We utilize the
pymoo library [4] (version 0.6.0) for GA. For each setting, we reserve a fraction of the total time budget for the PLS phase, with that fraction ranging from 0.1 to 0.5.

\begin{wrapfigure}{r}{0.35\linewidth}
    \vspace{-15pt}\includegraphics[width=\linewidth]{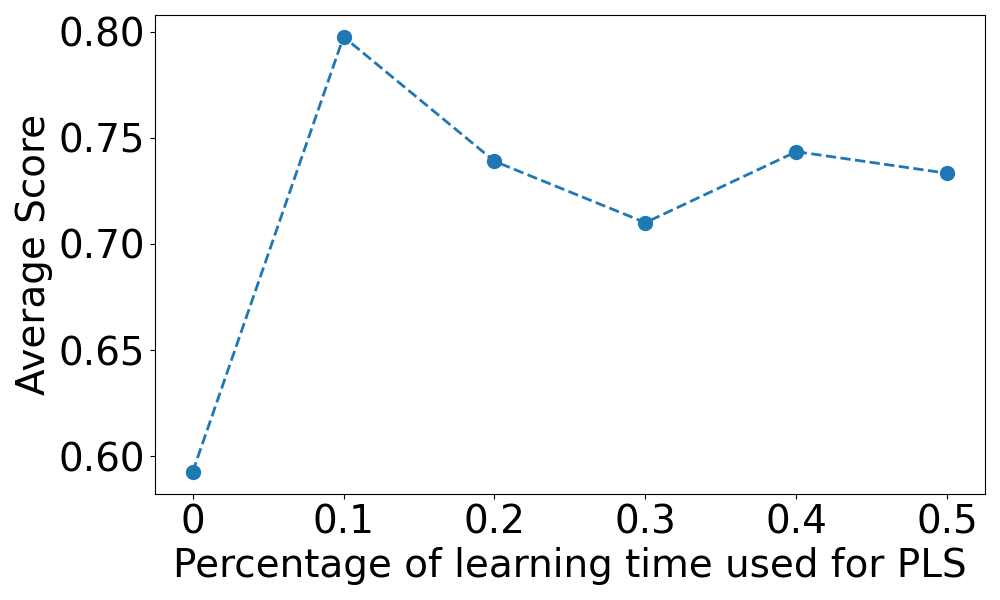}
    \caption{Mean score achieved by GA+MRE under different time allocations for the PLS phase.}
    \label{fig:selection_perf_time_pct}
    \vspace{-10pt}
\end{wrapfigure}

In Figure~\ref{fig:selection_perf_time_pct}, we report the mean score obtained when varying the fraction of time allocated to PLS. Here, $0$ indicates that PLS phase is not used and the optimization algorithm runs for the entire duration. We observe that allocating 10\% of the runtime to PLS yields the best performance, and use this in our subsequent tests of \sys. Thus, the most effective strategy is to devote the majority of time to exploration via the optimization algorithm, and then apply MRE briefly to more accurately select the best trace, rather than omitting MRE entirely or applying it for an extended duration.

\subsubsection{\textcolor{black}{Optimization Algorithm}}\label{sec:opt_alg}
We first investigate the effect of using the PLS phase with both GA and RG.
Note that all of the optimization methods can continue improving their scores with additional evaluations. Therefore, to ensure a fair comparison, we impose a fixed runtime budget. Specifically, we allow each optimizer to run for one hour per experiment.
When we enable the PLS phase, it consumes 10\% of this budget (i.e., 6 minutes), after which we record the best achieved score.

\begin{figure}[h]
    \centering
    \begin{subfigure}{0.48\linewidth}
        \centering
        \includegraphics[trim={0cm 0cm 0cm 0cm},width=0.7\linewidth]{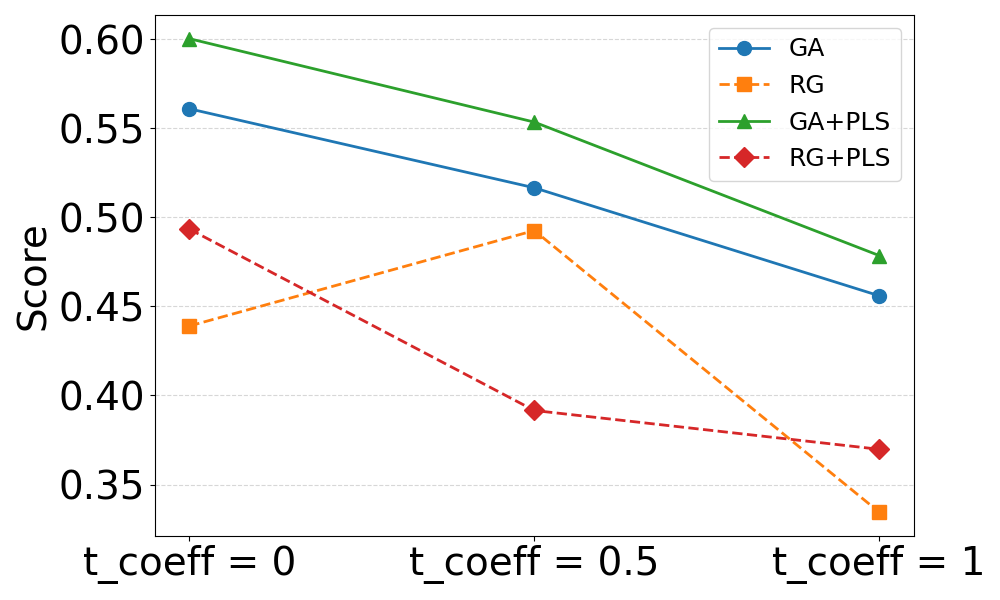}
        \caption{Performance of GA and RG with and without PLS.}
        \label{fig:RG_GA_selection}
    \end{subfigure}
    \hfill
    \begin{subfigure}{0.48\linewidth}
        \centering
        \includegraphics[trim={0cm 0cm 0cm 0cm},width=0.7\linewidth]{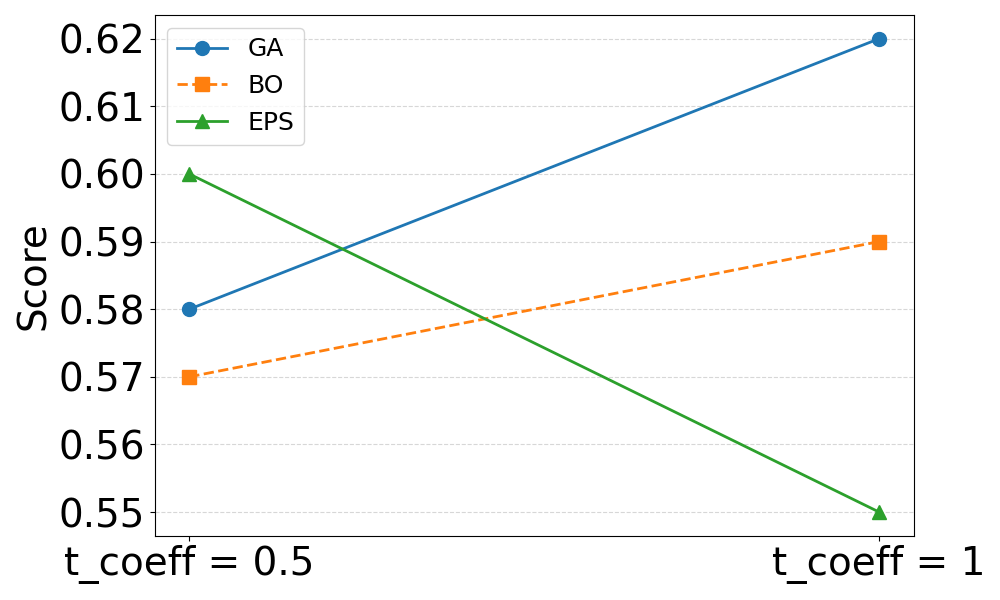}
        \caption{Performance of GA, BO \& EPS.}
        \label{fig:GA_BO_BL}
    \end{subfigure}
    \caption{Performance comparison of optimization algorithms and the impact of the PLS phase.}
    \vspace{-10pt}
    % \label{fig:sptcp_comp}
\end{figure}

In Figure~\ref{fig:RG_GA_selection}, we report the average \texttt{score}s obtained by GA and RG with and without the PLS phase across twenty scenarios, covering all pairwise combinations of \reference and \target drawn from the set \{\texttt{bbr}, \texttt{cubic}, \texttt{highspeed}, \texttt{reno}, \texttt{vegas}\} and using three values of $t_\textrm{coeff}$: 0, 0.5, and 1. Across all three scenarios, GA consistently outperforms RG, and the PLS phase consistently improves performance over runs without PLS. These results show that actively training an adversary to generate adversarial environments is more effective than relying on random generation. They also demonstrate that selecting from previously discovered environments using PLS for some amount of time at the ending is more effective than exploring for new adversarial environments.

We then compare the performance of GA, BO, and EPS. We implement BO using the \texttt{skopt} library (version 0.10.2). For EPS, we set the mutation probability ($\epsilon$) to 0.3 and use an elite set of size 10.
We evaluate them under the same combinations of \reference and \target as in the previous experiment, for two values of $t_{\textrm{coeff}}$ (0.5 and 1), over an extended runtime. Specifically, we run each experiment for two hours and allocate 10\% of the budget (12 minutes) to the PLS phase at the end.
Figure~\ref{fig:GA_BO_BL} shows the average performance of the three algorithms.
We observe no clear winner across the evaluated settings, but GA has slightly higher average performance and we use it in our remaining tests of \sys.
% \vspace{-5pt}

\subsubsection{Level of Parallelism} \label{sec:parallelism}
Next, we evaluate how the degree of parallelism affects performance. Recall (\S\ref{sec:execution-module}) that we use parallelism to run multiple emulation tests of the same trace.
We observe that parallel evaluation provides only marginal gains compared to a single evaluation per environment due to emulator interference, though we suspect this can be improved in the future (Appendix~\ref{appendix:parallelism}).
% \vspace{-10pt}
\subsection{\textcolor{black}{Comparison of TCP Protocols}}

So far we have evaluated the design choices within \sys. We now apply \sys to its intended purpose of testing protocols. Specifically, we evaluate the robustness of 17 TCP algorithms implemented in the Linux kernel.

\begin{figure}
    \centering
    \begin{subfigure}{0.48\linewidth}
        \centering
        \includegraphics[width=\linewidth, trim=3cm 3cm 4cm 0cm, clip]{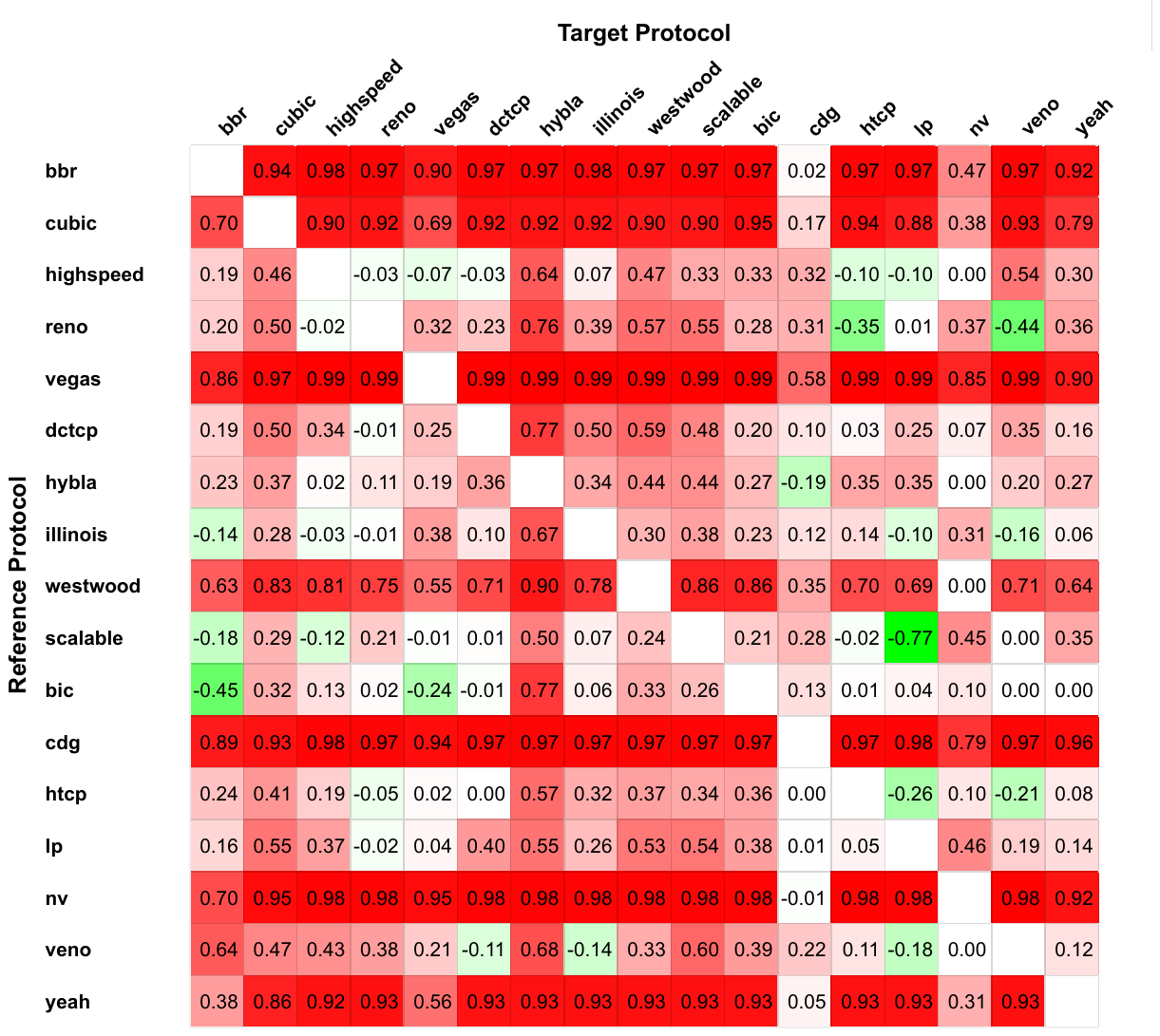}
        \caption{$t_\textrm{coeff}=0.5$}
    \end{subfigure}
    \hfill
    \begin{subfigure}{0.48\linewidth}
        \centering
        \includegraphics[width=\linewidth, trim=3cm 3cm 4cm 0cm, clip]{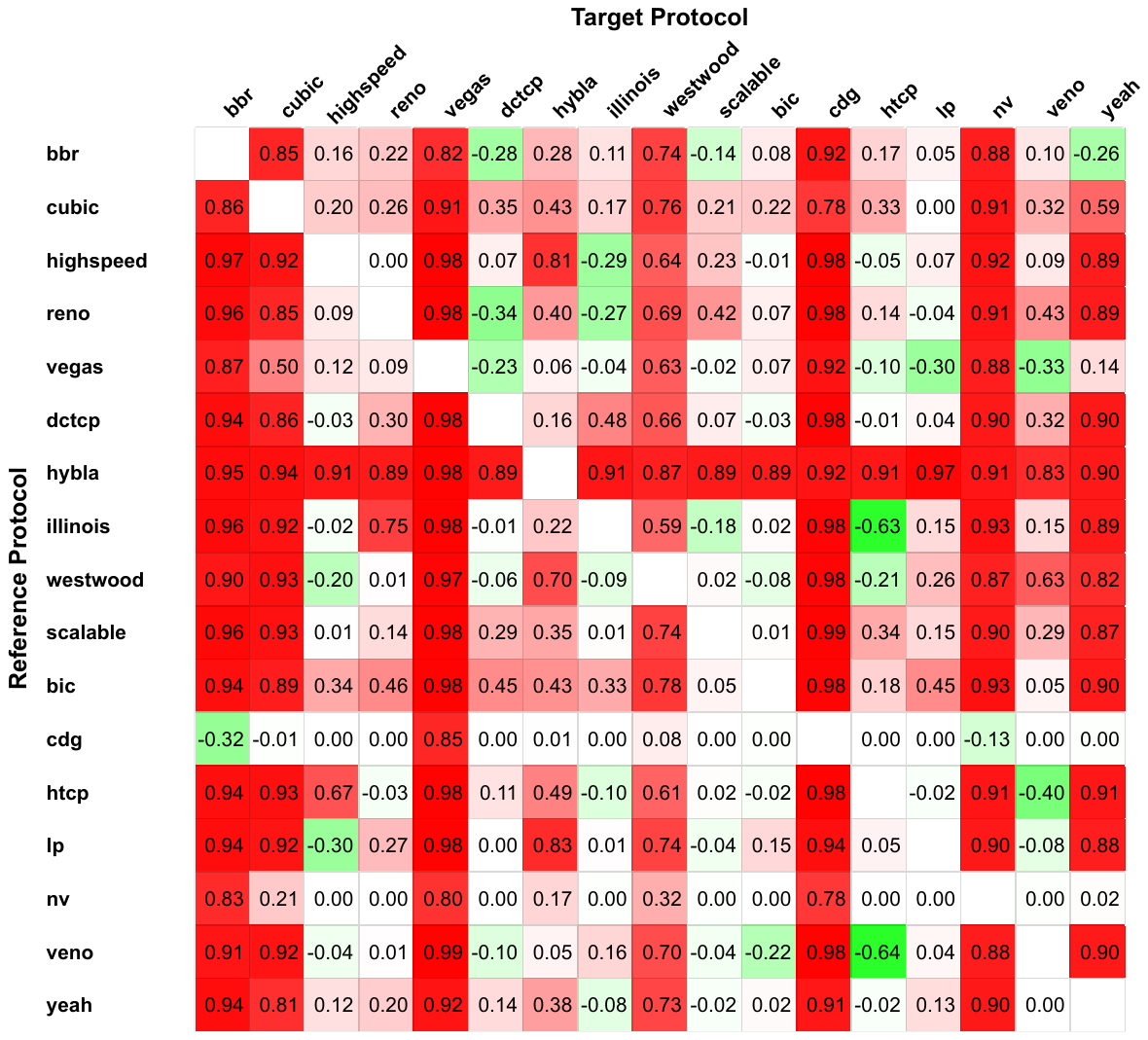}
        \caption{$t_\textrm{coeff}=1$}
    \end{subfigure}
    \caption{Pairwise robustness comparison of $17$ TCP protocols.}
    \label{fig:sptcp_comp}
\end{figure}

Using \sys, we conducted pairwise robustness comparisons of all 17 of the TCP protocols available in the Linux kernel. We gave \sys a time budget of 2 hour for each test, with the final 10\% of the time reserved for the selection phase. Figure \ref{fig:sptcp_comp} shows the results: each cell represents the relative performance difference between the corresponding \reference and \target in the trace that \sys discovered.

This result offers several notable insights. First, each column contains at least one substantially positive value, indicating that every protocol exhibits some vulnerability under appropriately chosen adversarial environments. Second, the ability to expose a particular protocol's weakness depends strongly on the choice of \reference. Not all protocols, when used as \reference, are capable of revealing the same vulnerability. For example, in the case of \texttt{bbr} with $t_{\textrm{coeff}}=0.5$, 6 out of the 16 protocols produce a large positive value, demonstrating that the effectiveness of adversarial discovery is reference-dependent.

% \begin{wrapfigure}{r}{0.5\linewidth}
%     \vspace{-5pt}
%     \includegraphics[width=\linewidth, trim=3cm 3cm 4cm 0cm, clip]{Figures/SPTCP_robustness_extended.pdf}
%     \caption{Pairwise robustness comparison of $17$ TCP protocols.}
%     \label{fig:sptcp_comp}
%     \vspace{-5pt}
% \end{wrapfigure}

\begin{wrapfigure}{r}{0.4\linewidth}
    \vspace{-20pt}\includegraphics[width=\linewidth]{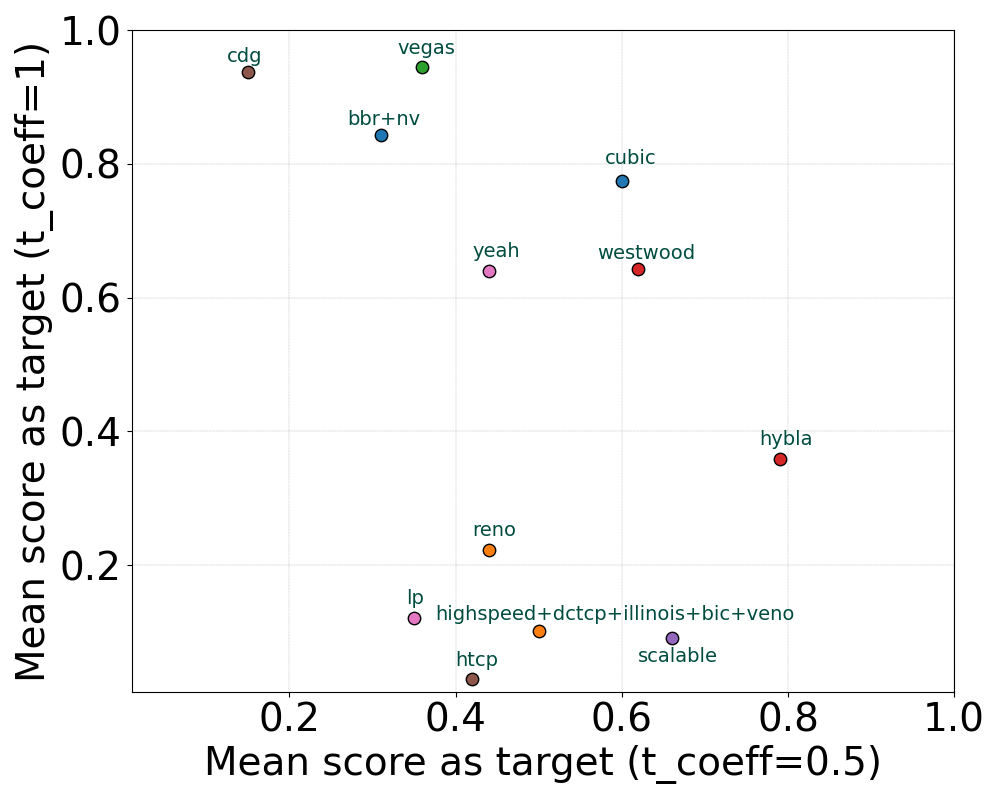}
    \caption{Mean \texttt{score} of each protocol as \target for $t_\textrm{coeff}=0.5$ and $t_\textrm{coeff}=1$.}
    \label{fig:sptcp_reg}
    \vspace{-20pt}
\end{wrapfigure}

In Figure~\ref{fig:sptcp_reg}, we plot each protocol's mean score as \target with $t_{\textrm{coeff}}=0.5$ on the x-axis and $t_{\textrm{coeff}}=1$ on the y-axis.
The mean score as \target measures \textbf{robustness}, i.e., lower values mean the protocol is harder to attack with adversarial environments. 

Interestingly, \texttt{cdg} achieves the lowest score along the x-axis, indicating that it offers the strongest robustness when balancing high throughput with low latency.
In contrast, \texttt{lp} and \texttt{htcp} achieve low scores along the y-axis, making them preferable when prioritizing high throughput without regard to latency.

% However, this observation should not be interpreted as a general conclusion. 
The relative position of each protocol in this plot is highly sensitive to the choice of $t_{\mathrm{coeff}}$, which determines the priority given to throughput versus latency. For example, when we set $t_{\mathrm{coeff}} = 1$ (i.e., assigning zero weight to latency), a specific bug in \texttt{bbr} is triggered (\S\ref{sec:bbr}), causing all scenarios with \texttt{bbr} as \target to yield scores above 0.8. Similarly, \texttt{vegas}, which relies on delay as a congestion signal, also becomes an easy \target under this setting, achieving scores above 0.9 regardless of the choice of \reference.
\vspace{-15pt}
\subsection{Looking Inside the Box}\label{sec:case}
In this section, we conduct a detailed analysis of the adversarial environments generated by \sys across different experiments. We aim to understand the underlying causes of adversarial behavior and, where possible, propose potential fixes to mitigate these issues.
% \vspace{-5pt}
\subsubsection{Case 1: \texttt{bbr}}\label{sec:bbr}
\sys successfully identified network environments where the Linux kernel implementation of \texttt{bbr} exhibited unexpected behavior. Upon examining the logs, we discovered an intriguing pattern in this anomalous behavior:
\sys manipulated the network conditions by maintaining a high RTT for about one second before abruptly reducing it. Some packets sent during the high RTT phase experienced long delays in reaching the destination, while later packets, benefiting from the lower RTT, arrived earlier. This out-of-order arrival triggered the receiver to generate duplicate acknowledgments (dup ACKs). Due to the low RTT, these dup ACKs quickly reached the sender. Upon receiving three dup ACKs, the sender initiated a fast retransmission for an earlier packet that was not lost but was simply taking longer to arrive.
Interestingly, despite the fast-retransmitted packet experiencing a lower RTT, the original delayed packet still arrived first. When the receiver received this packet, it sent an ACK for the next packet in sequence. However, upon later receiving the fast-retransmitted packet, an issue arose in its logic for updating the acknowledgment number.
As a result, even after the next expected packet was received, the acknowledgment number was not updated correctly. The receiver continued sending dup ACKs, ultimately forcing \texttt{bbr} into the retransmission timeout (RTO) phase.

To verify that TCP was indeed entering the RTO phase, we used eBPF to hook into the relevant TCP function. Additionally, we employed Wireshark to analyze whether the packet, for which the receiver continuously sent dup ACKs, was actually received by the kernel or dropped for some reason. Wireshark confirmed that the packet was indeed received by the kernel but was mishandled due to the specific sequence of events described earlier.

% One might question that the issue described above is not specific to \texttt{bbr}, then why doesn't this problem show up with other TCP protocols?
We observed that \texttt{bbr} behaves fundamentally differently from other TCP protocols. Unlike traditional congestion control algorithms, which send bursts of packets based on the congestion window and wait for their acknowledgments, \texttt{bbr} gradually sends packets while attempting to maintain a fixed number of packets in the network. This unique characteristic allowed \sys to manipulate network conditions in a way that induced the specific sequence of events with \texttt{bbr}, ultimately triggering the bug in the Linux kernel.

We further attempted to verify whether this behavior persists when using the latest Linux kernel version 6.13.7+ released by the BBR development team at Google. Under this version, \sys was unable to uncover scenarios exhibiting the previously observed bug. However, \sys did identify scenarios in which both \texttt{bbr} v1 and \texttt{bbr} v3 perform significantly worse than other TCP protocols such as \texttt{cubic} and \texttt{vegas}.
% Importantly, this behavior is not universal. \sys also discovers adversarial scenarios under the reversed configuration—when using \texttt{bbr }v1 or \texttt{bbr }v3 as \reference and \texttt{cubic} or \texttt{vegas} as \target—where \texttt{bbr} variants perform substantially better.

\begin{wrapfigure}{r}{0.35\linewidth}
    \vspace{-15pt}
    \centering
    \includegraphics[width=0.9\linewidth]{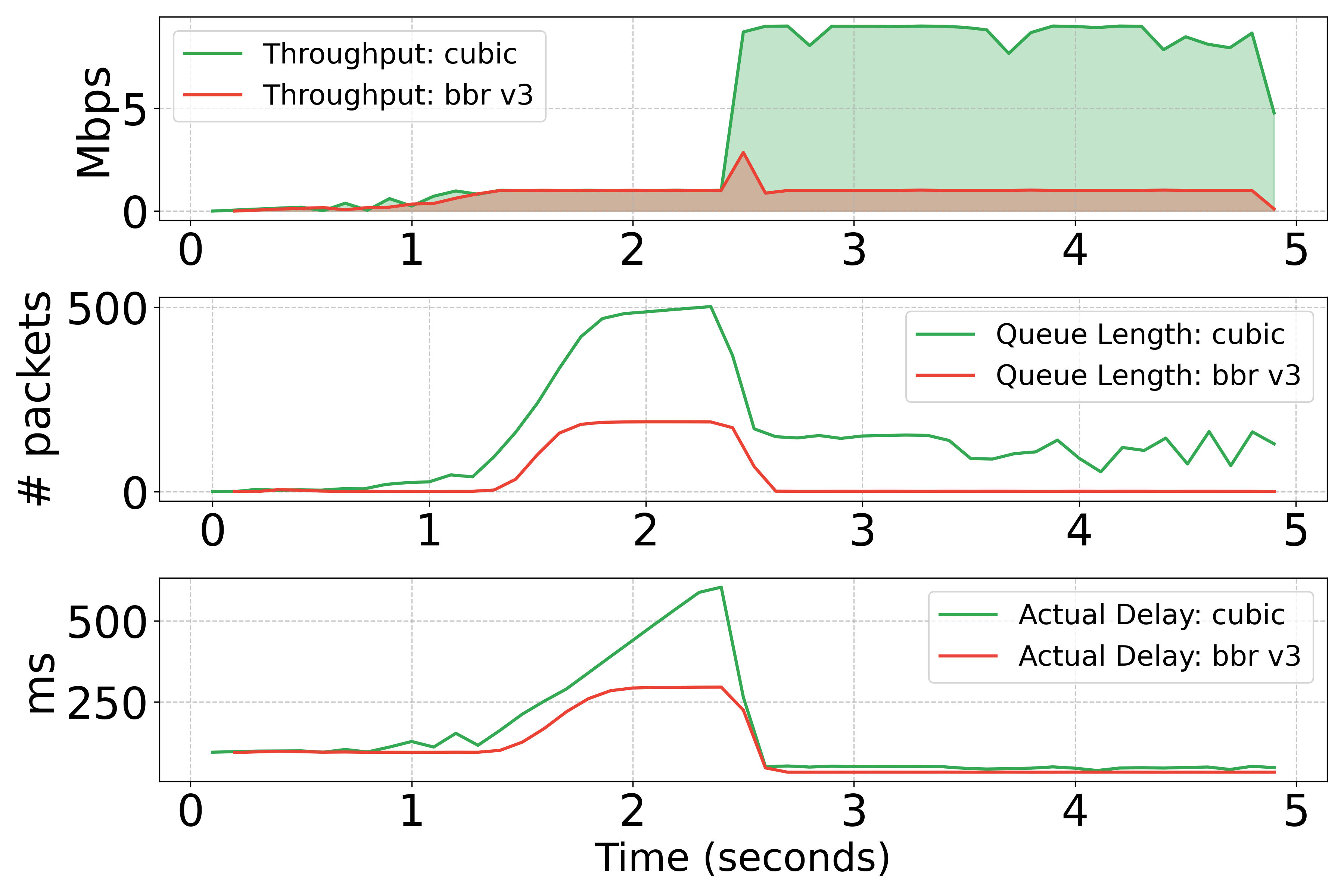}
    \caption{Adversarial environment for \texttt{bbr} v3.}
    \label{fig:cubic_vs_bbr_v3}
    \vspace{-15pt}
\end{wrapfigure}

Figure~\ref{fig:cubic_vs_bbr_v3} shows one such example. The throughput of \texttt{bbr} v3 remains low throughout the entire experiment, whereas \texttt{cubic} quickly ramps up and sustains a higher throughput. We extended the experiment duration to examine longer-term behavior and found that, over time, \texttt{bbr v3} gradually increases its throughput and eventually approaches higher rates. Upon further investigation, we observed that the BBR development team recommends using \texttt{bbr} with the \texttt{fq} qdisc, whereas Mahimahi defaults to \texttt{fq\_codel}. We modified our setup to use \texttt{fq}, but the same slow convergence behavior of \texttt{bbr} persisted.

% \vspace{-5pt}
\subsubsection{Case 2: \texttt{bbr} vs. \texttt{cubic}}
This scenario moves to a different objective -- fairness. We tasked \sys with identifying network environments in which the Linux kernel implementation of \texttt{cubic} is unfair to \texttt{bbr}. The goal of this experiment was to assess whether and to what extent \texttt{cubic} can be unfair to \texttt{bbr}, as the general perception in the networking community is that \texttt{bbr} tends to dominate other TCP flows \cite{ware2019modeling}. However, \sys discovered network conditions where \texttt{cubic} consumed nearly 100\% of the available bandwidth, leaving \texttt{bbr} starved. Interestingly, we observed a similar outcome when we reversed the roles of \reference and \target, instructing \sys to find scenarios where \texttt{bbr} is unfair to \texttt{cubic}.

Upon closer examination, we observed that \target initially operates as expected for a certain period (ranging from 40 to 80 seconds, depending on the environment).
However, beyond this point, it abruptly reduces its sending rate to a single packet at a time—waiting for an ACK before sending the next packet.
This behavior persisted for the remainder of the experiment (lasting between 25 and 60 seconds).
We reported this issue to the official \texttt{bbr} development team, who acknowledged the environment was problematic.
%the problem and informed us that they would provide a Linux kernel patch.
%This patch is expected to resolve the issue, allowing for a more accurate fairness comparison of Linux kernel TCP implementations.

% \vspace{-5pt}
\subsubsection{Case 3: DChannel}
We now move to a multipath test: are two paths always better than one? DChannel~\cite{sentosa2023dchannel} is designed to exploit a high bandwidth path (HBP) and a low latency path (LLP) in parallel. \sys was able to identify network environments where DChannel can perform three times worse in throughput compared to using only the high bandwidth path. 
DChannel steers packets along each path unbeknownst to the transport layer and application, which can cause the sender (i.e., CC) to misinterpret the observed network condition.  
It was reported previously in~\cite{touseef2023boosting} that DChannel can confuse delay-based TCP, such as \texttt{bbr}, as the increased delay caused by switching packet delivery from the LLP to the HBP path is misinterpreted as a congestion signal.  
However, their tests did not identify issues with DChannel when using \texttt{cubic}, a loss-based CC.

\sys discovered that DChannel performs poorly in a condition where there is a substantial difference in path latency between the HBP (120 ms one-way latency) and LLP (3 ms one-way latency).
We observed that \texttt{cubic} with DChannel experiences retransmission timeouts, leading to reduced throughput. 
This issue arises because DChannel's heuristic favors sending small control packets (e.g. TCP handshakes) over the LLP while sending large data packets to the HBP. 
This will confuse early RTT measurement~\cite{rfc6298}, as CC thinks that the path RTT is equal to the RTT of LLP, leading to a low RTO timer\footnote{We confirmed that the minimum RTO timer in the linux kernel is roughly 200 ms}. 
Therefore, when DChannel subsequently steers packets to the HBP which induces high latency, the RTO timer expires, triggering retransmission and causing CC to reduce its sending rate. 
We also confirmed that when we reduced HBP's RTT (to $<200$ ms, making it similar to what was used in \cite{touseef2023boosting}), DChannel outperforms the single path HB.

% We also tried to reduce the gap between the HB and LL path latency (making it similar to the case in \cite{touseef2023boosting}), and confirmed that we found the 
\begin{wrapfigure}{r}{0.35\linewidth}
\vspace{-5pt}
\includegraphics[width=0.9\linewidth]{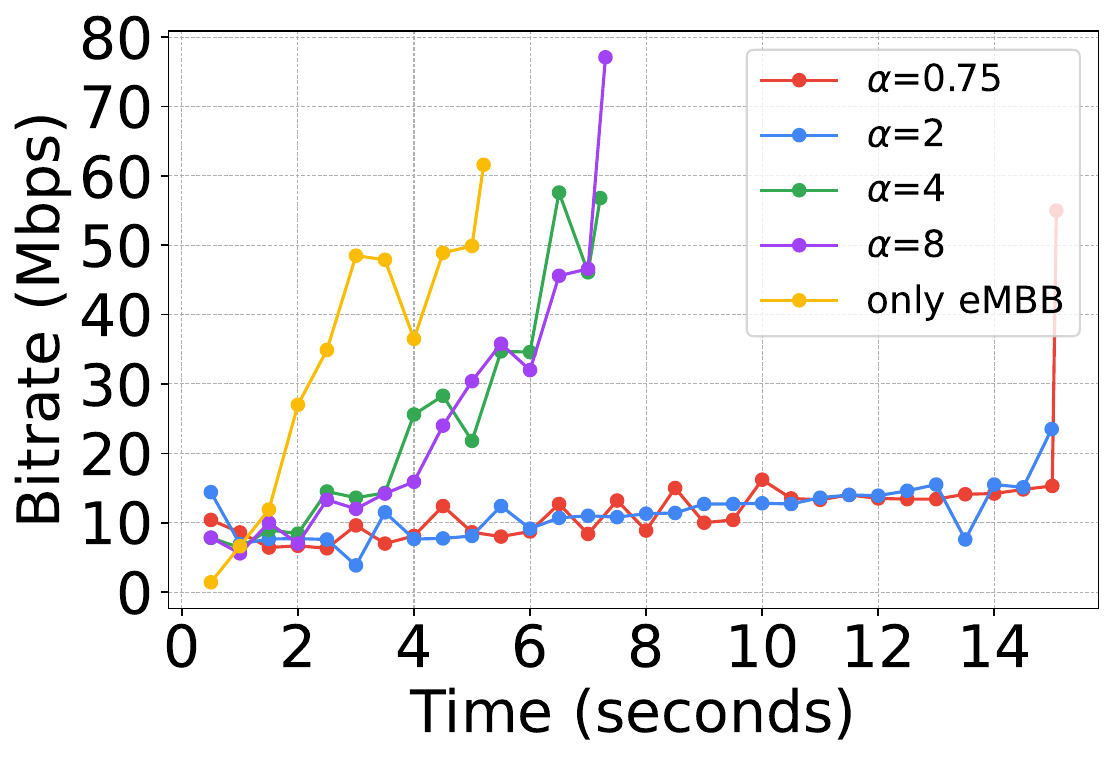}
  \caption{
  Bitrates across time for DChannel with different values of $\alpha$ and with only eMBB.
  } 
  \label{fig:dchannel_bitrates}
  \vspace{-10pt}
\end{wrapfigure}

We compared the throughput over time of DChannel (with different values of its internal $\alpha$ parameter) and a single HBP (labeled ``only eMBB'') in Figure \ref{fig:dchannel_bitrates} when we downloaded a 20MB file size. 
DChannel calculates the reward (R) of sending a packet to the LLP and its associated cost (C), and sends that packet to LLP only if $R > \alpha \times C$.
Thus, $\alpha$ directly impacts the number of packets steered to LLP.
DChannel sets the default $alpha$ value to $0.75$.
The result shows that \texttt{cubic} without DChannel (all-hb) outperforms \texttt{cubic} with DChannel regardless of the different $\alpha$ values, although increasing this value softens this issue.
This implies that DChannel's pathological behavior cannot be fixed by simply adjusting $\alpha$. 

% DChannel conducts a cost-reward analysis to determine whether to send packets to URLLC. If the reward exceeds $\alpha \times \text{cost}$, it selects URLLC; otherwise, it does not. Here, $\alpha$ is a tunable parameter, and through experimentation, the authors set its default value to $0.75$. As $\alpha$ increases, DChannel's behavior becomes more similar to using only eMBB.

% We plotted the bitrates over time in the network environment identified by \sys for DChannel with different values of $\alpha$ and when only eMBB is used. The default configuration lags significantly behind only eMBB. As we increase $\alpha$, DChannel's performance improves, but after reaching a certain point, it plateaus and cannot exceed or match the performance of only eMBB.
% \vspace{-5pt}
\subsubsection{Case 4: MPTCP with 1 link vs. 2 links}
We evaluated all four MPTCP congestion control protocols available in the Linux kernel—\texttt{lia}, \texttt{olia}, \texttt{balia}, and \texttt{wvegas}. In each case, \sys identified scenarios where performance degrades when the protocol is provided with two links compared to operating over a single link. We further investigated this behavior for \texttt{balia}, and observed that the throughput of \texttt{balia} with two links (\texttt{balia2}) was 85.5\% lower than that of \texttt{balia} with a single link (\texttt{balia1}).

\begin{wrapfigure}{r}{0.36\linewidth}
\vspace{-10pt}
  \includegraphics[trim={0cm 0cm 0cm 0cm},clip,width=\linewidth]{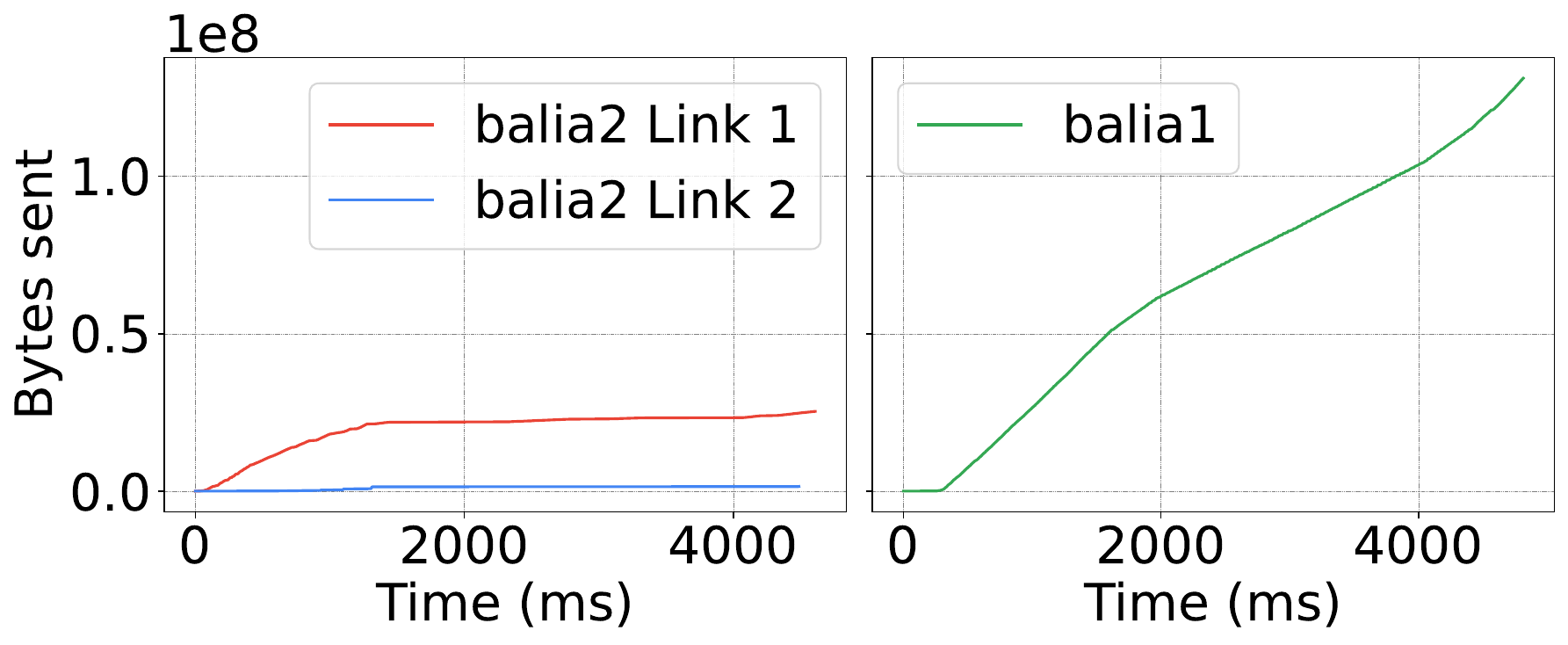}
  \caption{
  Number of bytes transmitted over time by \texttt{balia1} and \texttt{balia2}.
  } 
  \label{fig:balia_seqs}
  \vspace{-20pt}
\end{wrapfigure}

Figure \ref{fig:balia_seqs} illustrates the number of bytes sent over time for the two scenarios.
Initially, \texttt{balia2} performs better by taking advantage of the high bandwidth of link 1.
However, \texttt{balia1} quickly catches up and surpasses \texttt{balia2} after approximately 500 ms.
At around 1200 ms, an interesting event occurs: \texttt{balia2} reduces its sending rate on both link 1 and link 2, despite link 1 still having high bandwidth.
We discovered that the receiver had delayed acknowledging the receipt of a packet by roughly 40 ms.
This delay led \texttt{balia2} to significantly lower its sending rate across both links.

% \vspace{-5pt}
\subsubsection{Case 5: \texttt{bbr v1} vs. \texttt{bbr v3}}
Here, we analyze the fairness behavior of \texttt{bbr v1} and \texttt{bbr v3} using the latest official kernel (version~6.13.7+) released by the BBR development team.
For this experiment, we first allow \sys to manipulate bandwidth and latency parameters, and then continue execution for an additional one minute with fixed parameters to allow the protocols to stabilize. We then measure performance during the final ten seconds of the run. Interestingly, \sys identifies scenarios in which \texttt{cubic} achieves 12 times more throughput than \texttt{bbr v1} when run concurrently. However, when \texttt{bbr v1} is replaced with \texttt{bbr v3}, the roles reverse: \texttt{cubic} obtains 11.6 times less throughput compared to \texttt{bbr v3}\footnote{The anomalous behavior observed here is a direct consequence of the suboptimal behavior of the \texttt{bbr} versions discussed earlier}.

% However, this observation does not necessarily imply that, under this adversarial environment, \texttt{bbr v3} is more unfair to \texttt{cubic} than \texttt{bbr v1}. The low throughput exhibited by \texttt{bbr v1} stems from its previously discussed suboptimal behavior rather than genuine unfairness. A meaningful fairness comparison can only be conducted once these suboptimal behaviors in the TCP protocols are addressed.

% \vspace{-10pt}
\section{\textcolor{black}{Discussion and Future Work}}
\label{sec:future-work}

We see several directions for extending \sys beyond the current setting.

First, while we demonstrate \sys in the context of congestion control, the core idea of adversarial environment generation applies more broadly to systems whose performance depends on complex and dynamic environments. Examples include adaptive bitrate (ABR) video streaming, microservice workloads, and cloud resource management. In fact, we have already built a prototype integration of \sys for ABR video. Extending \sys to new domains, however, introduces domain-specific challenges, particularly in defining the environment space and designing suitable adversarial generators. For instance, ABR systems require modeling video chunk sizes and player behavior, while microservices require capturing workload patterns and inter-service dependencies. We expect each domain to require customized adversary designs and evaluation metrics.

Second, \sys can be extended to support richer and more realistic network environments. Our current implementation, based on Mahimahi, models time-varying bandwidth and latency, with packet loss arising implicitly from buffer constraints. However, congestion control performance also depends on other factors such as workload characteristics (e.g., flow size distributions, concurrency, and burstiness), network topology, and deployment scale. Incorporating these factors would require extending both the environment representation and the underlying emulator or simulator. For example, modeling network topology would require representing environments as graphs, which in turn would necessitate adapting the optimization algorithms to operate over structured inputs. Similarly, incorporating workload dynamics would require expanding AdvNet to generate not only link conditions but also traffic patterns.

Third, our current framework does not attempt to explain why a given adversarial environment degrades protocol performance. \sys focuses on generating challenging scenarios, leaving root-cause analysis to the user, as in many existing test-generation systems. An interesting direction for future work is to automatically characterize and summarize the regions of the environment space that induce adversarial behavior, which could provide actionable insights for protocol designers.

Overall, while \sys provides a first step toward adversarial environment generation for protocol design, extending it to broader settings will require rethinking both the environment representation and the adversary design to account for domain-specific complexities.

%It would be interesting to investigate whether the randomly generated adversarial scenarios, derived from the patterns extracted by \sys, could be leveraged to develop enhanced versions of the protocol under test, similar to the concept of GANs, where the generated environments are used to retrain and improve the discriminator.

% \vspace{-10pt}
\section{Related Work}
\label{sec:related}
\textbf{Testing Network Protocols.} 
Testing the performance of network protocols has long been an important part of networking.
% because applications using
% the network rely on high performance to meet key quality metrics like latency, rebuffering and download time. 
As a result, a vast collection of simulators~\cite{issariyakul2009introduction, riley2010ns} and emulators~\cite{white2002integrated, netravali2015mahimahi} provide a means to test many network protocols.

There are also other works aiming at creating more robust protocols.
For instance, Pantheon~\cite{yan2018pantheon} runs a variety of congestion control protocols on real-world paths and emulated paths intended to reflect a spectrum of real world scenarios.
Puffer~\cite{yan2020learning} streams video using different adaptive video streaming protocols on real-world network. 
Though they run on real-world paths, however, both Pantheon and Puffer may not be as appropriate for identifying and replicating the exact conditions that cause undesirable behavior, since such conditions might only arise when other uncontrolled real world processes take place. 
% In addition, when undesirable behavior is observed, it may not be as easy to understand what part of a trace was important for inducing the bad behavior.
The most basic difference, however, is that \cite{yan2018pantheon,yan2020learning} and similar systems test real-world conditions, but do not seek out (potentially unknown) worst-case conditions.
In general, our work should not be seen as a replacement for running realistic tests. 
Instead, realistic tests and adversarial tests are complementary.
Our work should be regarded as a way to generate a broader set of test cases to improve robustness and understand a protocol’s flaws.

\textbf{Adversarial Testing of Networking Protocols.} 
Our work builds on an earlier workshop paper~\cite{gilad2019robustifying} that employs RL to identify adversarial instances where ABR and CC protocols significantly deviate from optimality, using a model of optimality.
The present paper develops a system for emulation rather than simulation 
% incorporates a reference performance measurement as an effective way to define a benchmark 
and specializes this for three use cases.
% , develops a GA approach, includes user-configurable environmental parameter bounds, tests a much wider array of real-life protocol implementations using emulation rather than simulation, and modifies the emulator to support multiple links and provide useful information for debugging. 
Perhaps most importantly, we performed an extensive set of tests of real protocols and root-caused problems, resulting in discovery and understanding of multiple Linux kernel and other issues.

More recently, another workshop paper~\cite{ray2022cc} harnessed GA to generate adversarial network environments and traffic patterns for several CC algorithms, successfully uncovering bugs in the NS3 implementation of \texttt{bbr} and \texttt{cubic}.
In contrast, our work is designed for emulation rather than only simulation, and deals with associated noise tolerance using PLS; defines network environments 
% more intuitively and understandably 
as time-varying bandwidths and latencies (as opposed to PDOs), which we showed results in better performance.

GENET~\cite{xia2022genet} introduced the idea of finding performance gaps between two protocols, for a different goal than ours, namely incorporating curriculum learning to improve the training effectiveness of RL-based network protocols.

% MetaOpt \cite{namyar2024finding} proposed a bi-level optimization approach to find performance gaps between a heuristic with either another heuristic or the optimal algorithm. However, it requires the heuristic to be converted to a convex optimization problem and the heuristic would have to search solutions from a collection of constraints.
% While TCP algorithms can be transformed into convex optimization problems \cite{anderson2003stability}, this approach imposes several challenges: it necessitates assumptions about the algorithm’s behavior, demands significant effort from protocol designers to reformulate new implementations as convex optimization problems, and, most critically, limits the ability to test real-world implementations that are or will be deployed for practical use.

\cite{arun2021toward} developed a Congestion Control Anxiety Controller to formally verify certain properties of congestion control (CC) algorithms, while \cite{arun2022starvation} introduced a theoretical model for analyzing two competing flows and explored the potential starvation of delay-convergent CCs.
The key distinction between \sys and these works is that \sys focuses on real \emph{implementations}, whereas these studies aim to validate specific characteristics of CC \emph{algorithms in simplified network models}. 
% This leads to significant differences in approaches, of learning-based adversarial environment generation (\sys) vs. formal verification of a model (\cite{arun2021toward}). \cite{arun2021toward}'s method would not meet our needs because it would be extremely difficult to formally verify the real implementations we target.
We believe these directions are both valuable and complementary.

\textbf{ML-based input generation.}
ML-based input generation has been explored previously; we discuss this in more detail in \S\ref{sec:ml_rel}.

\textbf{\textcolor{black}{Formal methods–based adversarial testing.}}
Formal methods have also been used to systematically generate adversarial inputs for complex systems. 
% \cut{Prior work has applied techniques such as model checking, symbolic execution, and constraint solving to explore corner cases and uncover violations of protocol specifications.} 
For example, LTEInspector \cite{hussain2018lteinspector} leverages symbolic analysis to generate adversarial test cases for 4G LTE control-plane protocols. Symbolic execution engines such as \cite{cadar2008klee} and \cite{godefroid2005dart} have been widely used to automatically explore program paths and generate inputs that trigger errors. Similarly, model checking tools like \cite{holzmann1997model} have enabled systematic exploration of protocol state spaces to detect correctness violations.  
% \cut{More broadly, formal approaches enable exhaustive or guided exploration of well-defined state spaces, providing strong guarantees about coverage and correctness.} 
However, these techniques typically require precise protocol models or specifications and often struggle to scale to large, continuous, or poorly structured input spaces such as network environments with time-varying dynamics. 
% \cut{In contrast, our approach does not rely on formal specifications; instead, it learns to generate adversarial environments through iterative interaction with the system under test, making it more suitable for complex and high-dimensional settings.}

% \vspace{-14pt}
\section{Conclusion}
In this paper, we seek to identify implementation flaws or suboptimal performance within transport protocols.
We achieve this by employing a learning algorithm that acts as an adversary, generating adversarial network environments attempting to maximize the performance gap between \target and \reference.
\textcolor{black}{Our framework has successfully conducted robustness testing on all TCP and MPTCP protocols implemented in the kernel. \sys gave an understanding of overall protocol robustness,  identified several issues and provided insight into the causes.}

% \fix{Reminders / To Do:}

\section{\textcolor{black}{Acknowledgments}}

% Identification of funding sources and other support, and thanks to
% individuals and groups that assisted in the research and the
% preparation of the work should be included in an acknowledgment
% section, which is placed just before the reference section in your
% document.

% This section has a special environment:
% \begin{verbatim}
%   \begin{acks}
%   ...
%   \end{acks}
% \end{verbatim}
% so that the information contained therein can be more easily collected
% during the article metadata extraction phase, and to ensure
% consistency in the spelling of the section heading.

% Authors should not prepare this section as a numbered or unnumbered {\verb|\section|}; please use the ``{\verb|acks|}'' environment.
This project was supported by NSF awards 2008971 \& 2326576, ISF grants 1508/17 \& 3481/24 and BSF grants 2019798 \& 2023663.
\bibliographystyle{ACM-Reference-Format}

% \appendix

\appendix
\section{Level of Parallelism}
\label{appendix:parallelism}
Before determining the optimal level of parallelism, we first investigate the maximum degree of parallelism that the underlying machine can reliably support. To quantify the overhead introduced by evaluating environments in parallel rather than sequentially, we conduct a controlled experiment. Specifically, we fix the network environment at 1~Gbps bandwidth and 5~ms one-way delay, and execute both \texttt{bbr} and \texttt{cubic} ten times each while varying the level of parallelism from 1 to 10 concurrent evaluations.

\begin{figure}[h!]
    \centering
    \resizebox{0.7\linewidth}{!}{%
    \begin{minipage}{\linewidth}
        \centering
        \begin{subfigure}[t]{0.49\linewidth}
            \includegraphics[width=\textwidth]{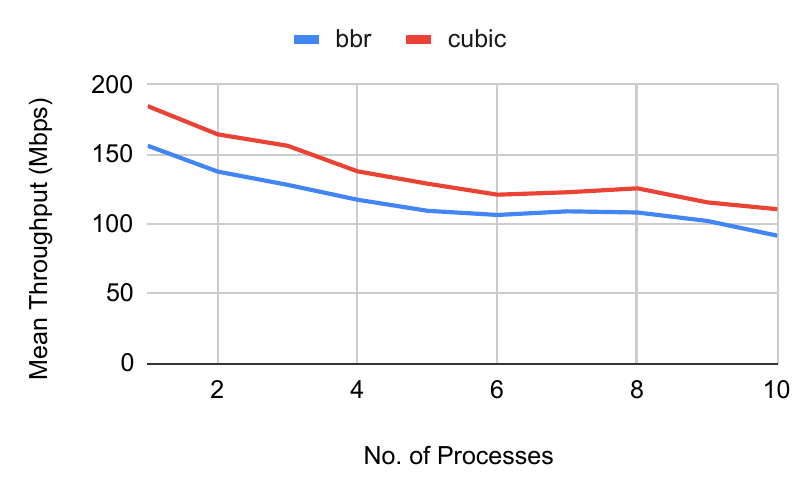}
        \end{subfigure}
        \begin{subfigure}[t]{0.49\linewidth}
            \includegraphics[width=\textwidth]{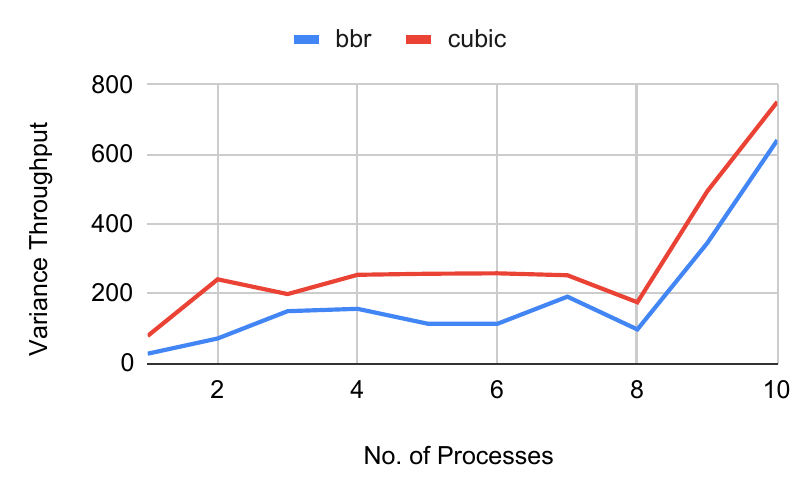}
        \end{subfigure}

        \begin{subfigure}[t]{0.49\linewidth}
            \includegraphics[width=\textwidth]{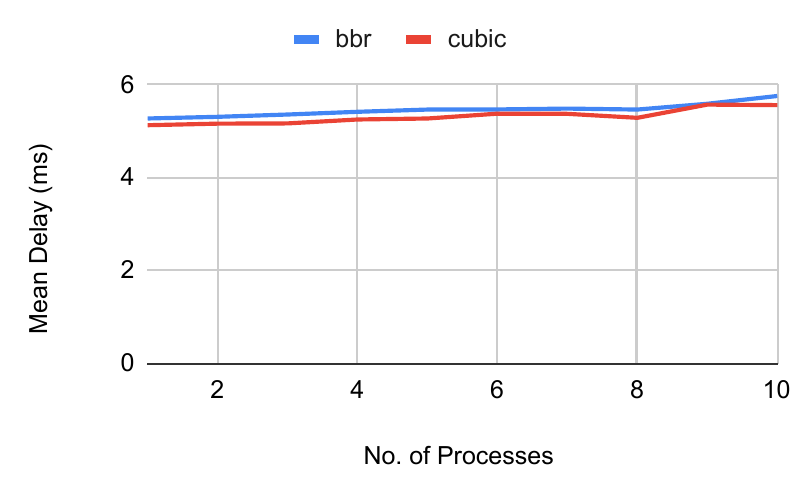}
        \end{subfigure}
        \begin{subfigure}[t]{0.49\linewidth}
            \includegraphics[width=\textwidth]{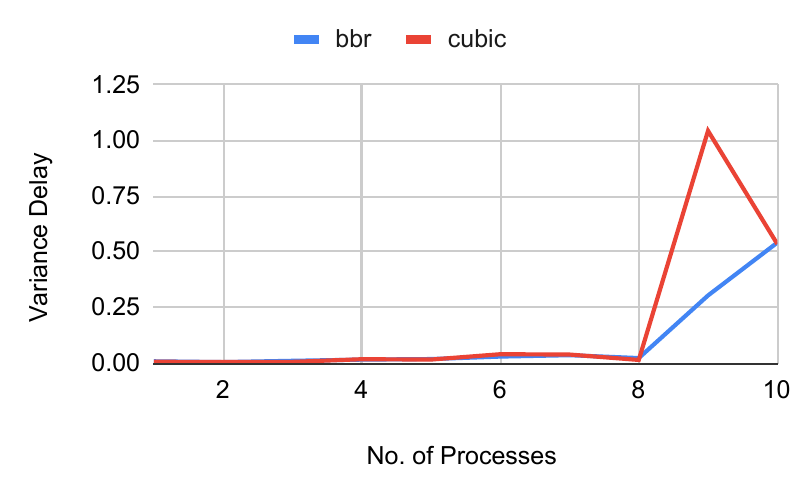}
        \end{subfigure}
    \end{minipage}%
    }
    \caption{The effect of number of parallel executions on experienced throughput and delay.}
    \label{fig:parallelism_overhead}
\end{figure}

In Figure~\ref{fig:parallelism_overhead}, we report the mean and variance of throughput and latency obtained by \texttt{bbr} and \texttt{cubic} when varying the parallelism level from 1 to 10 simultaneous evaluations. While the effect on mean latency is negligible, the mean throughput gradually decreases as parallelism increases. Beyond 8 parallel evaluations (i.e., at 9 and 10), the variance increases sharply. This behavior arises from hardware contention: the experimental server had 30 CPU cores, and each execution required three dedicated cores––one for the emulator, one for the \texttt{iperf} server, and one for the \texttt{iperf} client. With 9 or more parallel executions, the system begins to oversubscribe CPU resources, leading to higher variability in performance. Based on this observation, we constrained the parallelism level to at most 8 in our experiments.

For finding out the correct level of parallelism, we run \sys for 250 iterations (each iteration translates to picking a network environment and computing its \texttt{score}) using \texttt{cubic} as \reference and \texttt{bbr}, \texttt{highspeed}, \texttt{reno}, and \texttt{vegas} as \target. For each target protocol, we vary the number of parallel evaluations from 1 to an upper bound of 8.

\begin{wrapfigure}{r}{0.35\linewidth}
\includegraphics[width=0.95\linewidth]{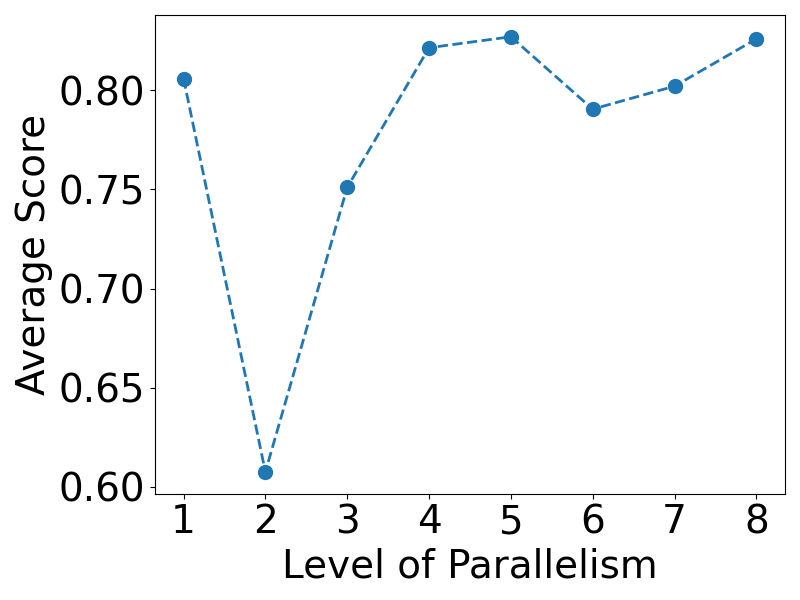}
        \caption{Effect of parallelism on achieved score by GA.}
        \label{fig:level_of_parallelism}
\end{wrapfigure}

Figure~\ref{fig:level_of_parallelism} shows the average result across the four \target protocols. A single evaluation per environment yields a strong baseline score. As parallelism increases, score decreases and then increases again. This is due to two effects. First, we found that protocol throughput in the emulator degrades noticeably (Figure \ref{fig:parallelism_overhead}), even moving from 1 to 2 parallel tests, and even with the precaution of running on separate cores as noted earlier. This means \sys is effectively receiving somewhat inaccurate information about trace performance (note that we always perform final evaluation of the score of its output using a single emulation at a time). Second, as parallelism increases, the greater number of trials per trace gives \sys more samples to combat noise. The shape of these two competing effects is such that 5-way parallelism produces the best results for a given wall-clock time.

The difference in performance between using 1 and 5 evaluations per environment is relatively small, and the former would be preferable when aiming to minimize CPU usage. However, since the overall wall-clock time remains unchanged, we adopt 5 evaluations per environment for the remaining experiments to obtain the highest possible score. If a user chooses to use 1 evaluation per environment, however, it would not significantly affect the results.

Regarding the problem of parallelism degrading emulation performance, our optimizations of Mahimahi (e.g., generating PDOs on the fly rather than as a large input) reduced but did not eliminate the  degradation. We suspect this is likely due to a bottleneck that could be improved, perhaps by running emulations in separate VMs. In addition, it would be possible to explore other uses of parallelism such as testing different traces in parallel rather than multiple runs of the same trace. We leave these optimizations to future work.

\section{Post-Learning Selection Algorithms}
\label{appendix:selection_algorithms}
Due to the prohibitively high runtime of evaluating selection strategies using real-world protocols, we performed a controlled simulation study. 

In particular, in each trial of the simulation, we generate 50 Gaussian random variables. The true mean of each variable is selected randomly at the beginning of the simulation, uniformly between 0 and 100. Then, we apply one of the selection algorithms to identify the variable with the highest mean. Each selection algorithm was given the same sampling budget, representing the total number of samples it was allowed to draw from the random variables.

\begin{wrapfigure}{r}{0.5\linewidth}
    \vspace{-10pt}
    \centering
    \includegraphics[width=\linewidth]{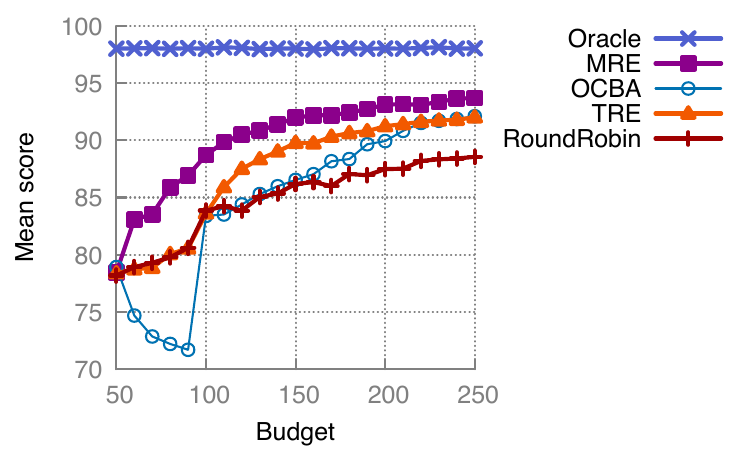}
    \caption{Mean score achieved by different selection algorithms for different budgets.}
    \label{fig:selection_comparison}
    \vspace{-5pt}
\end{wrapfigure}

We test the following algorithms. \textbf{Oracle} returns the true optimal random variable, i.e., that with maximum mean; since there are 50 random variables with means chosen uniformly at random from 0-100, this score averages just below 100. \textbf{RoundRobin} splits the budget equally among all variables (plus 1 extra for some, if the budget is not a multiple of 50). \textbf{OCBA} is the algorithm of~\cite{478499}, and \textbf{MRE} is our algorithm; these two are described in \S\ref{sec:pls}. \textbf{Two-Round Elimination (TRE)} is a simpler two-round evaluation that begins by evaluating all environments a fixed number of times (2); then, the top 25\% of performers from the first round are repeatedly re-evaluated until the evaluation budget is exhausted.

Figure~\ref{fig:selection_comparison} shows the average results over 2000 independent trials, where each trial consists of generating a fresh set of 50 random variables and allowing the selection algorithm to identify the best one. The score of a run is the true mean of the variable selected by the algorithm. MRE outperforms all competing strategies, leading us to adopt MRE as the selection mechanism in \sys's PLS phase.

For OCBA, we observe that when the sampling budget is below 100, the mean score remains very low and even decreases as the budget increases. This behavior arises because OCBA relies heavily on accurate variance estimates for each alternative. With 50 variables in the pool, a budget below 100 allows at most one sample per variable during the warm-up phase, yielding no meaningful variance estimates. Once the budget reaches 100, OCBA’s performance improves rapidly, and the mean score increases steadily with additional budget. As the budget approaches 250, OCBA becomes the second-best performing method among the evaluated selection strategies.

%\textbf{(1) PickFirst:} This baseline method always selects the first environment, without performing any comparison.

%\textbf{(2) RoundRobin:} In this method, the environments are evaluated in a round-robin fashion.

%\textbf{(3) Two-Round Elimination (TRE):} This strategy begins by evaluating all environments a fixed number of times. The top 25\% performers from this initial phase are then extracted out and repeatedly re-evaluated until the evaluation budget is exhausted.

%\textbf{(4) Multi-Round Elimination (MRE):} This algorithm extends TRE by employing successive elimination rounds. Each environment is evaluated once, and the environments are sorted by their mean performance. In each subsequent stage, the lower-ranked half of the population is discarded, continuing until at most five environments remain. At that point, the remaining environments are repeatedly evaluated without further elimination.

%\textbf{(5) Optimal Computing Budget Allocation (OCBA) \cite{478499}:} OCBA allocates evaluations adaptively by prioritizing environments exhibiting higher uncertainty or those that are close in score to the current best. Thus, environments with smaller observed performance gaps from the best or larger empirical variances receive more evaluation attempts, leading to statistically efficient identification of the best environment.

\section{DChannel Parameter Bounds}\label{appendix:dchannel}

\begin{table*}[h!]
    \centering
    {\small
    \begin{tabular}{c|c|c|c|c|c|c|c|}
    \cline{2-8}
    % &\multicolumn{7}{c|}{\textbf{DChannel}} \\ \cline{2-8}
    &\multicolumn{2}{c|}{\textbf{eMBB}}&\multicolumn{2}{c|}{\textbf{URLLC}}&\multicolumn{3}{c|}{\textbf{eMBB \& URLLC}}\\ \cline{2-8}
    &\textbf{Bandwidth}&\textbf{Latency}&\textbf{Bandwidth}&\textbf{Latency}&\textbf{Duration}&\textbf{Queue Length}&\textbf{Data size}\\ \cline{1-8}
    \multicolumn{1}{|c|}{\textbf{Lower Bound}}&15 Mbps&10 ms&2.5 Mbps&2 ms&50 ms&100&500 KB\\ \cline{1-8}
    \multicolumn{1}{|c|}{\textbf{Upper Bound}}&150 Mbps&125 ms&5 Mbps&5 ms&50 ms&5000&50000 KB\\ \cline{1-8}
    \end{tabular}
    }
    \caption{Parameters used in all experiments with \sys for DChannel.
    The lower and upper bounds for each parameter are defined per timestep.
    We adopted these values from the original paper, with the exception of adjusting the bounds for \emph{URLLC latency} and \emph{Data size} to increase the likelihood of challenging DChannel.
    % Therefore, if \sys is searching for network traces with 2 timesteps, the lower and upper bounds for each of the two eMBB bandwidths in each trace will be 15 Mbps and 150 Mbps, respectively.
    }
    \label{tab:params_dc}
    \vspace{-10pt}
\end{table*}
\vspace{-10pt}
\section{ML-based input generation}
\label{sec:ml_rel}
Using ML to generate complex test input is not new.
An automatic test-data generation tool utilizing GA was proposed in \cite{pargas1999test}.
For instance, recent work integrating ML and test input generation has been done with fuzz testing for PDF parsers~\cite{godefroid2017learn}. 
Using ML to generate tests for software dates back more than a decade~\cite{shu2007testing}.
Our approach bears similarities to generative adversarial networks (GANs)~\cite{goodfellow2014generative}. 
GANs, however, represent a supervised learning approach and, more importantly, our goal is substantially different. 
GANs are typically useful for generating new data that is indistinguishable from an existing dataset (see~\cite{briand2008novel} for a recent example of applying GANs to networking
domains). 
In our context, we do not already possess traces of challenging network conditions to which GANs might be applied, but aim to create such challenging traces.

\end{document}